%% file: paper-extended.tex
\documentclass[acmsmall,screen]{acmart}
\usepackage{cmll}
\usepackage{mathpartir}

\newcommand{\tmRec}{\mathrm{rec}}
\newcommand{\tyNat}{\mathrm{Nat}}
\newcommand{\conZero}{\mathsf{zero}}
\newcommand{\conSucc}{\mathsf{succ}}
\newcommand{\conRefl}{\mathsf{refl}}

\newcommand{\Let}{\mathrm{let}}
\newcommand{\LetPair}{\mathrm{letpair}}
\newcommand{\In}{\mathrm{in}}
\newcommand{\If}{\mathrm{if}}
\newcommand{\Then}{\mathrm{then}}
\newcommand{\Else}{\mathrm{else}}
\newcommand{\dupNat}{\mathrm{dupNat}}
\newcommand{\Match}{\mathrm{match}}

\newcommand{\istype}{\mathrm{type}}
\newcommand{\isctxt}{\mathrm{ctxt}}

\newcommand{\BoolTy}{\mathrm{Bool}}
\newcommand{\cTrue}{\mathsf{true}}
\newcommand{\cFalse}{\mathsf{false}}
\newcommand{\ListTy}{\mathrm{List}}
\newcommand{\cNil}{\mathsf{nil}}
\newcommand{\cCons}{\mathsf{cons}}

\newcommand{\Rtype}{\mathbf{R}}
\newcommand{\rIntro}{\mathbf{R}}
\newcommand{\rElim}{\mathbf{R}^{-1}}

\newcommand{\natinf}{\mathbb{N}_{-\infty}}

\newcommand{\Ty}{\mathrm{Ty}}
\newcommand{\RTm}{\mathrm{RTm}}
\newcommand{\Tm}{\mathrm{Tm}}

\newcommand{\Set}{\mathrm{Set}}

\newcommand{\cat}[1]{\mathcal{#1}}
\newcommand{\op}{\mathsf{op}}

\newcommand{\ConsFree}{\emph{Cons-free}}

\newcommand{\LinPreorder}{\mathrm{LinPreorder}}

\newcommand{\lemref}[1]{\hyperref[#1]{Lemma \ref*{#1}}}
\newcommand{\exref}[1]{\hyperref[#1]{Example \ref*{#1}}}
\newcommand{\judgeref}[1]{\hyperref[#1]{Judgement \ref*{#1}}}
\newcommand{\propref}[1]{\hyperref[#1]{Proposition \ref*{#1}}}
\newcommand{\diagref}[1]{\hyperref[#1]{Diagram \ref*{#1}}}
\newcommand{\thmref}[1]{\hyperref[#1]{Theorem \ref*{#1}}}
\newcommand{\defref}[1]{\hyperref[#1]{Definition \ref*{#1}}}

\newcommand{\AgdaModule}[1]{\textcolor{blue}{\tt #1}}

\setcopyright{rightsretained}
\acmDOI{10.1145/3632918}
\acmYear{2024}
\copyrightyear{2024}
\acmSubmissionID{popl24main-p453-p}
\acmJournal{PACMPL}
\acmVolume{8}
\acmNumber{POPL}
\acmArticle{76}
\acmMonth{1}
\received{2023-07-11}
\received[accepted]{2023-11-07}

\begin{document}

\title{Polynomial Time and Dependent Types}
\subtitle{Extended Version}

\author{Robert Atkey}
\email{robert.atkey@strath.ac.uk}
\orcid{0000-0002-4414-5047}
\affiliation{%
  \institution{University of Strathclyde}
  \streetaddress{26 Richmond Street}
  \city{Glasgow}
  \country{UK}
  \postcode{G1 1XH}
}


\begin{abstract}
  We combine dependent types with linear type systems that soundly and
  completely capture polynomial time computation. We explore two
  systems for capturing polynomial time: one system that disallows
  construction of iterable data, and one, based on the LFPL system of
  Martin Hofmann, that controls construction via a payment
  method. Both of these are extended to full dependent types via
  Quantitative Type Theory, allowing for arbitrary computation in
  types alongside guaranteed polynomial time computation in terms. We
  prove the soundness of the systems using a realisability technique
  due to Dal Lago and Hofmann.

  Our long-term goal is to combine the extensional reasoning of type
  theory with intensional reasoning about the resources intrinsically
  consumed by programs. This paper is a step along this path, which we
  hope will lead both to practical systems for reasoning about
  programs' resource usage, and to theoretical use as a form of
  \emph{synthetic computational complexity theory}.
\end{abstract}



\maketitle

\section{Introduction}
\label{sec:introduction}

Type Theory is often claimed to be ideal for Computer Science,
combining programming and proof in one unifying system, so a happy
programmer can verify while they program, and program while they
verify. From a broader Computer Science view, however, Type Theory
lacks the ability to talk about the very thing that makes Computer
Science interesting -- the fact that computation is everywhere bounded
by the resources in time and space that we can afford to give it.

Typically, Type Theory only speaks of the public face that programs
present to the world -- if you input things like this, you get things
that look like that -- but cannot bring itself to mention the true
cost of programs' execution. One can encode costs by embedding another
programming language in Type Theory, for example \cite{GueneauCP18},
or one can synthesise costs by treating resource counting as a
computational effect, for example \cite{Danielsson08,NiuSGH22}, but
neither of these capture the intrinsic costs of the programs we write
in Type Theory. These techniques deliver only \emph{conspicuous
  consumption}, not speaking of the real resources consumed.

In this paper, we propose a method for extending dependent Type Theory
with a means for constraining the intrinsic computational complexity
of programs written in the theory. We concentrate on linear type
systems that soundly and completely capture polynomial time
computation, the commonly used standard for feasible resource usage,
and extend these systems to dependent types. The additional
expressivity of dependent types allows us use these characterisations
of polytime to further functionally characterise the classes of
non-deterministic and bounded-error probabilistic polynomial time.

We use techniques from \emph{Implicit Computational Complexity}
theory, which provides intrinsic characterisations of complexity
classes in terms of logical systems or programming languages. We
review the techniques that we use in \autoref{sec:linear-intro}. To
adapt these systems to dependent types, we use \emph{Quantitative Type
  Theory} (QTT) \cite{atkey18qtt,mcbride16}, a combination of linear
and dependent types. We review QTT in \autoref{sec:qtt}.

Our long-term goal is to combine the extensional reasoning of Type
Theory with intensional reasoning about the resources intrinsically
consumed by programs. This paper is a first step along this path,
which we hope will lead both to practical systems for reasoning about
programs' resource usage as well as their extensional behaviour, and
to theoretical use as a form of \emph{synthetic computational
  complexity theory}. We discuss these possibilities further in
\autoref{sec:conclusion}.

\subsection{Contributions and Content}

This paper makes the following contributions to the theory and use of
linear dependent type theory and implicit computational complexity:
\begin{enumerate}
\item We formulate two systems that combine linear type theory for
  polytime computation with full dependent types, using Quantitative
  Type Theory. The systems are presented in \autoref{sec:qtt}. The
  linear typing discipline required for enforcing polytime is provided
  by QTT, but we also need to carefully add constructs for
  non-iterable datatypes (\autoref{sec:noniter-qtt}) and the two kinds
  of natural number iterator that we consider
  (\autoref{sec:cons-free-qtt} and \autoref{sec:lfpl-qtt}). Porting
  the natural number iterators from the simply typed to the
  dependently typed setting requires careful annotation of the rules
  to ensure that the correct information is available for type
  checking, while also not allowing too much information to be made
  available at runtime that would violate the polytime soundness
  property. A further contribution of this paper is the addition of
  reflection types to QTT, \autoref{sec:qtt-reflection}, which allow
  statements about polytime realisability to be reflected into
  types.
\item We demonstrate the utility of the combination of polytime and
  dependent types in \autoref{sec:programming-polytime}. Just as in
  the simply typed world, we have an expressive language for writing
  polytime programs. With the additional power of dependent types, we
  can also prove properties of these programs. A simple example is
  proving that a polytime sorting program actually sorts. Using QTT reflection,
  we can go further and represent the class of polytime problems, with
  polytime reductions between them, as dependent pairs
  (\autoref{sec:polytime-problems}). Our final examples use dependent
  types to give monadic presentations of the complexity classes of
  Non-deterministic Polynomial (NP) time and Bounded-error
  Probabilistic Polynomial (BPP) time. Since these classes rely on
  specific semantic correctness criteria, it is not possible to capture them in
  a simply typed system for polytime.
\item We prove the polytime soundness of our systems via a
  realisability argument in \autoref{sec:soundness} and
  \autoref{sec:realising-iteration}. Our construction is an extension
  of the amortised complexity realisability constructions of
  \citet{dallago11realisability}. We extend their work to our
  dependently typed setting, and also give a realisability
  interpretation of datatypes directly, instead of via second-order
  impredicative encodings. The technical content of these sections has
  been formalised in the Agda proof assistant
  \cite{norell2008dependently}, and is included in the associated
  artefact \cite{atkey2023polydep-artefact}.
\end{enumerate}

Before we get to the contributions above, we present, in
\autoref{sec:linear-intro}, two linear simply typed systems for
polytime, adaptations of systems already present in the
literature. Our paper concludes with a discussion of further related
work and the outlook for future work in \autoref{sec:conclusion}.

\section{Affine Linear Typing and Polytime}
\label{sec:linear-intro}

Not long after Girard introduced Linear Logic \cite{girard87}, it was
observed that its resource sensitivity could be turned to describing
computational complexity classes by purely logical means. Typically, a
logical system is described for which the process of reducing a proof
to a normal form (often by cut elimination) is guaranteed to always be
accomplished within a certain complexity bound. Moreover, the system
is usually proven to be complete for the relevant complexity class by
constructing a simulation of some known representation. Such systems
that characterise polytime include Bounded Linear Logic (BLL), which
uses explicit polynomials in the formulas \cite{bll92} and Soft Affine
Logic (SAL) \cite{Lafont04}, which does not explicitly represent time
information in formulas, but uses a restricted form of Linear Logic's
$\oc$ modality instead. Light Linear Logic (LLL) \cite{lll98} is
another ``counting-free'' system for polytime.

Viewing logical systems though the Curry-Howard correspondence, the
idea arises that one could define functional programming languages
that characterise complexity classes such as polytime. SAL has been
transformed into a programming language by \citet{BaillotM04}, and
likewise for LLL by \citet{BaillotGM10}. \citet{hofmann99lfpl}
proposed a new programming language, Linear Functional Programming
Language (LFPL), that uses a novel ``payment'' system to track
iteration.

There are at least two ways that a functional programming language can
be seen as representing polynomial time, differing in how the size of
the problem to be computed is measured. One approach is to consider
closed expressions, combining the program with its input, and
computation of the result is polynomial time in the combined size. A
second approach is that the input is ``externally'' provided, where we
consider open terms with a free variable representing the input. So a
judgement $x : \tyNat \vdash M : A$ declares a program that computes
results of type $A$ in time polynomial in the size of the natural
number $x$. We take this latter approach in this paper.

With a view to extending to dependent types in \autoref{sec:qtt}, we
take an approach slightly different to much of the polytime linear
logic literature. We use explicit datatypes and eliminators, rather
than using impredicative encodings via universal types. We are closer
to Hofmann's original LFPL (though not a later presentation of it by
\citet{dallago11realisability}) than BLL, SAL or LLL.

In this section, we review the use of linear types to capture polytime
by presenting two systems, one based on ideas from SAL and the second
more explicitly based on LFPL.

\subsection{Affine Linear $\lambda$-Calculus}

For this section, the affine linear $\lambda$-calculus we will use
will have linear functions and $\otimes$-products. Contexts are
treated up to permutation of entries, so uses of exchange are
implicit.
\begin{mathpar}
  \inferrule*
  { }
  {\Gamma, x : A \vdash x : A}

  \inferrule*
  {\Gamma, x : A \vdash M : B}
  {\Gamma \vdash \lambda x.M : A \multimap B}

  \inferrule*
  {\Gamma_1 \vdash M : A \multimap B \\
    \Gamma_2 \vdash N : A}
  {\Gamma_1, \Gamma_2 \vdash M\,N : B}

  \inferrule*
  {\Gamma_1 \vdash M : A \\ \Gamma_2 \vdash N : B}
  {\Gamma_1, \Gamma_2 \vdash (M,N) : A \otimes B}

  \inferrule*
  {\Gamma_1 \vdash M : A \otimes B \\ \Gamma_2, x : A, y : B \vdash N : C}
  {\Gamma_1, \Gamma_2 \vdash \Let\,(x,y)=M\,\In\,N : C}
\end{mathpar}
These rules are standard, so we do not describe them further except to
note how affine linear typing uses presence or absences in a context
to control resource usage. If a variable is in the context it must be
used at most once (variables that are not used are absorbed by the
additional context in the variable rule). The fact that this
discipline interferes with dependent types is one of the reasons we
turn to QTT when we wish to add dependent types in \autoref{sec:qtt}.

\subsection{No Recursion, Only Case Analysis}
\label{sec:linear-case-analysis}

It is not too difficult to see that reduction of linear
$\lambda$-terms always takes a number of steps linearly proportional
to the size of the term. This is because every $\beta$-redex
substitutes each term into at most one variable, reducing the size of
the term by one each time.

We can increase the expressivity, but not the computational
complexity, of the system by adding datatypes that do not allow
iteration. These can be used for representation but not for driving
computation. We include the rules here to show how linearity must be
preserved in these rules and to foreshadow their dependently typed
counterparts in \autoref{sec:noniter-qtt}. The first type is the
booleans, which are non-recursive and so would not allow
iteration anyway:
\begin{mathpar}
  \inferrule*
  { }
  {\vdash \cTrue, \cFalse : \BoolTy}

  \inferrule*
  {\Gamma_1 \vdash M : \BoolTy \\ \Gamma_2 \vdash N_1 : A \\ \Gamma_2 \vdash N_2 : A}
  {\Gamma_1, \Gamma_2 \vdash \If\,M\,\Then\,N_1\,\Else\,N_2 : A}
\end{mathpar}
The if-then-else rule is careful to ensure that the resources used by
the eliminated $\BoolTy$ and the resources used by the chosen branch
are accounted for separately. The two branches must have the same
resource usage.

Construction and case analysis of lists are given by the following
rules:
\begin{mathpar}
  \inferrule*
  { }
  {\vdash \cNil : \ListTy(A)}

  \inferrule*
  {\Gamma_1 \vdash M : A \\ \Gamma_2 \vdash N : \ListTy(A)}
  {\Gamma_1, \Gamma_2 \vdash \cCons(M,N) : \ListTy(A)}

  \inferrule*
  {\Gamma_1 \vdash M : \ListTy(A) \\
    \Gamma_2 \vdash N_1 : B \\
    \Gamma_2, h : A, t : \ListTy(A) \vdash N_2 : B}
  {\Gamma_1, \Gamma_2 \vdash \Match\,M\,\{\cNil \mapsto N_1; \cCons(h,t) \mapsto N_2\} : B}
\end{mathpar}
We can construct lists arbitrarily but only do case analysis on
them. If we wish to explore a list to a arbitrary depth it must be
driven by a type we can iterate over.

With booleans and lists, we can construct several other useful
types. For example, to simulate Turing machines, one can construct a
$\mathrm{Tape}$ type as a Zipper (\citet{Huet97})
$\ListTy(\BoolTy) \otimes \BoolTy \otimes \ListTy(\BoolTy)$,
representing a position on the tape with the items before, under, and
after the head.

\subsection{The Cons-free System}
\label{sec:cons-free-intro}

Polynomial time is usually seen as a proxy for ``feasible''
computation. On the face of it, there does not seem to be any
particular reason why polynomials have anything to do with
feasibility. However, one can arrive at the definition of polynomial
time in three steps, by assuming that (i) iterating over the whole
input is feasible; (ii) if two computations are feasible, then so is
their composition; and (iii) performing a feasible computation for
every element of the input is also feasible. It is the last point that
allows complexities of arbitrary polynomial degree to be constructed
(we will see this in action in the completeness construction below and
soundness proofs in \autoref{sec:realising-iteration}).

Following these ideas, let us assume that the input is a natural
number, so we assume that there is some type of natural numbers
$\tyNat$. For point (i), we must be able to iterate over these natural
numbers, so we use a linear iterator defined by this typing rule:
\begin{displaymath}
  \inferrule*
  {\vdash M_z : A \\ x : A \vdash M_s : A \\ \Gamma \vdash N : \tyNat}
  {\Gamma \vdash \tmRec\,N\,\{\conZero \mapsto M_z; \conSucc(x) \mapsto M_s\} : A}
\end{displaymath}
Note that in the zero, $M_z$, and sucessor, $M_s$, cases, the context
is empty to ensure that these cases may be invoked as many times as
required. Point (ii) above is automatically satisfied by being in a
typed $\lambda$-calculus, where it is difficult to stop functions from
being composable. For point (iii), the iterator as given does not
allow us to nest iterations. Once the natural number input $n$ has
been used for an iteration, the linear typing discipine prevents us
from using it again (note the two separate contexts $\Gamma_1$,
$\Gamma_2$ in the rule for application). In order to allow nested
iterations, we add an operator to duplicate numbers:
\begin{displaymath}
  \inferrule*
  {\Gamma \vdash M : \tyNat}
  {\Gamma \vdash \dupNat\,M : \tyNat \otimes \tyNat}
\end{displaymath}
Somewhat surprisingly, this system is now sound and complete for
polynomial time. Crucially, this depends on the two things we have
\emph{not} allowed. First, we have disallowed the construction of new
natural numbers via the $\conZero$ and $\conSucc$
constructors\footnote{Actually, $\conZero$ would be acceptable, as
  well as any constant natural number. It is only unrestricted use of
  $\conSucc$ that is dangerous.}. If we were to permit this, then we
could use iteration over the input to construct addition,
multiplication (by repeated addition), and then exponentials (by
repeated multiplication). We therefore refer to this system as the
\ConsFree{} system.  Because we cannot construct values of type
$\tyNat$ within the system, complete programs in this system are open
terms as we explained at the start of this section.

The second prohibited feature is the ability to duplicate values of
function type, even though we have allowed duplication of iterable
naturals. If we were to allow this, then we would be able to sneak in
a form of constructors for natural numbers by encoding them as
eliminators that duplicate a function for every $\conSucc$ step.

We will see in \autoref{sec:consfree-sound} that this system is sound
for polytime by a realisability argument. Completeness can be seen
more directly by constructing a function that iterates a function for
a statically known polynomial number of times in the size of the
input. Assume that we have a known polynomial
$p(n) = c_dn^d + \dots + c_0$ of degree $d$ with natural number
coefficients and some single step function
$f : \mathit{St} \multimap \mathit{St}$ over a state type
$\mathit{St}$ that runs to completion for input of size $n$ in $p(n)$
steps. Then, using the iterator above we can iterate $f$ over a
$\tyNat$ representing the size of the input:
\begin{displaymath}
  \begin{array}{l}
    I_1 : \tyNat \multimap \mathit{St} \multimap \mathit{St} \\
    I_1 = \lambda n. \lambda s.\tmRec\,n\,\{\conZero \mapsto s; \conSucc(s) \mapsto f\,s \}
  \end{array}
\end{displaymath}
To achieve higher degrees, we can use $\dupNat$ to nest iterations:
\begin{displaymath}
  \begin{array}{l}
    I_{k+1} : \tyNat \multimap \mathit{St} \multimap \mathit{St} \\
    I_{k+1} = \lambda n. \lambda s.
    \begin{array}[t]{@{}l}
      \Let\,(n,n') = \dupNat\,n\,\In\\
      \tmRec\,n\,\{\conZero \mapsto s; \conSucc(s) \mapsto I_k\,n'\,s \}
    \end{array}
  \end{array}
\end{displaymath}
By further use of $\dupNat$ and composition to handle addition of
polynomials, the function $f$ can now be iterated $p(n)$ many times,
where $n$ is the input $\tyNat$. Thus, the \ConsFree{} system can
represent all polytime computations.

\subsection{Diamond Trading with LFPL}
\label{sec:lfpl-intro}

The \ConsFree{} system is sound and complete for polytime, but is quite
awkward from the point of view of functional programming. It allows us
to iterate over natural numbers that come from the input but does not
allow us to build further values to do iteration on. For example, if
our input is a list, then we cannot transform it into a binary search
tree and then flatten it, we must always refer back to the original
natural number input. Even dividing the input into two halves to be
treated separately is difficult.

A more flexible system was proposed by \citet{hofmann99lfpl}. Instead
of completely prohibiting construction of data, the \emph{Linear
  Functional Programming Language} (LFPL) allows construction if it is
paid for by values of type $\Diamond$ (``diamonds''):
\begin{mathpar}
  \inferrule*
  {\Gamma \vdash M : \Diamond}
  {\Gamma \vdash \conZero(M) : \tyNat}

  \inferrule*
  {\Gamma_1 \vdash M : \Diamond \\ \Gamma_2 \vdash N : \tyNat}
  {\Gamma_1, \Gamma_2 \vdash \conSucc(M, N) : \tyNat}
\end{mathpar}
To construct a $\conZero$, we must have a $\Diamond$ to pay for it,
and likewise, to construct a $\conSucc$ we must pay a $\Diamond$. We
can think of $\Diamond$s as an unit of iterable data. Iterability is
``saved up'' in data during construction, and released during
iteration. Diamonds cannot be created by a program itself, for the
same reason that constructors were prohibited in the \ConsFree{}
system, but they are released from iterable data during iteration. The
LFPL natural number iterator has the following typing rule:
\begin{displaymath}
  \inferrule*
  {d : \Diamond \vdash M_z : A \\ d : \Diamond, x : A \vdash M_s : A \\ \Gamma \vdash N : \tyNat}
  {\Gamma \vdash \tmRec\,N\,\{\conZero(d) \mapsto M_z; \conSucc(d,x) \mapsto M_s \} : A}
\end{displaymath}
The difference with the \ConsFree{} iterator above is that the
$\conZero$ and $\conSucc$ cases now both have an additional binding of
type $\Diamond$. This allows some form nesting of iterations: during
an iteration over the input, the program can accumulate $\Diamond$s to
use for iteration over substructures that are smaller than the current
point in the iteration. A construction, due to \citet{AehligS02},
illustrates how this leads to all polytime computations. As above, we
assume a polynomial $p(n)$ and a step function
$f : \mathit{St} \multimap \mathit{St}$ that needs to be iterated
$p(n)$ times. We construct a linear iterator:
\begin{displaymath}
  \begin{array}{l}
    I_1 : (\tyNat \otimes \mathit{St}) \multimap (\tyNat \otimes \mathit{St}) \\
    I_1 = \lambda (n, s).\,\tmRec\,n\,\{
    \begin{array}[t]{@{}lcl}
      \conZero(d)&\mapsto&(\conZero(d), s);\\
      \conSucc(d,(n,s)) &\mapsto& (\conSucc(d,n),f\,s) \}
    \end{array}
  \end{array}
\end{displaymath}
Note that this iterator returns the natural number input as well as
the new state. LFPL does not allow duplication of iterable inputs, so
we must always reconstruct it if we want to do further
iteration. Addition of polynomials is accomplished by composition of
iterators. To raise the degree, we again use a nesting iterator:
\begin{displaymath}
  \begin{array}{l}
    I_{k+1} : (\tyNat \otimes \mathit{St}) \multimap (\tyNat \otimes \mathit{St}) \\
    I_{k+1} = \lambda (n,s).\,\tmRec\,n\,\{
    \begin{array}[t]{@{}lcl}
      \conZero(d)&\mapsto&(\conZero(d),s); \\
      \conSucc(d,(n,s)) &\mapsto& \Let\,(n,s)=I_k\,(n,s)\,\In\,(\conSucc(d,n), s) \}
    \end{array}
  \end{array}
\end{displaymath}
Unlike in the \ConsFree{} system, this iterator does not raise
the degree of the nested iterator directly. Rather, the iterator $I_k$
on the input $n$ performs $\binom{n}{k}$ iterations. As observed by
Aehlig and Schwictenberg, this is sufficient because the binomials
form a basis for the vector space of all polynomials.

Despite this slightly more involved completeness construction, the
advantage of LFPL is that it is now easy to have arbitrary iterable
datatypes and to transform between them. We need only take the
introduction and elimination rules for any inductive datatype and add
$\Diamond$ premises to the introduction rules and $\Diamond$ bindings
to the eliminators.

\section{Polytime Quantitative Type Theory}
\label{sec:qtt}

We have now seen the \ConsFree{} and LFPL systems for
capturing polytime by means of linear typing and restricted
iteration. We now look to extend these systems to include dependent
types by building upon \emph{Quantitative Type Theory} (QTT)
\cite{atkey18qtt,mcbride16}. This section reviews QTT and describes
how we have adapted it to the polytime systems we saw in the previous
section.

\subsection{Quantitative Type Theory}
\label{sec:qtt-basic}

Integrating linear and dependent types is not straightforward due to
the conflict between the linear typing discipline regarding presence
of a variable as only bestowing the right to use it once, and the
dependent typing regime that uses variables both in types (for
specification purposes) and in terms (for computational purposes),
syntactically yielding multiple uses of the same variable.

QTT is a system that resolves this conflict by recording usage of
variables with annotations from a semiring. It sits in the general
area of systems that use semiring annotations to measure resource usage
\cite{BrunelGMZ14,GhicaS14,OrchardLE19}. The key feature of QTT, an
insight owing to \citet{mcbride16}, is that usage of variables in
types counts for $0$-usage in terms of the semiring used. This allows
us to use normal type theory as a specification language, while also
enjoying the benefits of linear typing for programs. The term typing
judgement of QTT has the following form:
\begin{displaymath}
  x_1 \stackrel{\rho_1}: S_1, \dots, x_n \stackrel{\rho_n}: S_n \vdash M \stackrel\sigma: T
\end{displaymath}
where the annotations $\rho_i$ are all from the semiring being
used. The annotation $\sigma$ is either $0$ or $1$, indicating whether
we are in the erased ($\sigma = 0$) fragment, where all the normal
rules of type theory apply, or the in the non-erased (``present'',
``realisable'', $\sigma = 1$) fragment, where a restricted typing
discipline applies.  As we shall see below in the cases of
$\Sigma$-types, iterable types, and LFPL's $\Diamond$ type, the
separation of QTT into these two fragments allows an expressive
combination of reasoning using full type theory with the benefits of
linear typing.

In the remainder of this sub-section, we describe the core of QTT. As
well as the term typing judgement given above, QTT also has judgements
for well-formed contexts ($\Gamma~\isctxt$) and types
($\Gamma \vdash T~\istype$), and definitional equality of types and
terms ($\Gamma \vdash S \equiv T~\istype$ and
$\Gamma \vdash M \equiv N \stackrel\sigma: S$). It is an invariant of
the system that types are always well-formed in a context with all
annotations $0$, i.e., $\Gamma \vdash S~\istype$ implies
$0\Gamma = \Gamma$. An important admissible rule of the system, along
with weakening and substitution, is that of $0$-ing:
\begin{displaymath}
  \inferrule*
  {\Gamma \vdash M \stackrel1: S}
  {0\Gamma \vdash M \stackrel0: S}
\end{displaymath}
This rule allows us to take any term $M$ in the $\sigma = 1$ fragment
and treat it as if it were in the $\sigma = 0$ fragment, and hence use
it for specification purposes in types. As we add novel rules to QTT
in the following sections, we will be careful to maintain the
admissibility of this rule.

In this section, we give an overview of the rules of QTT. The full
rules, including all equality rules, are presented in
\autoref{sec:rules}.

\subsubsection{Natural-number Usages}

We use an instantiation of QTT with the natural number semiring, with
the usual semiring structure of addition and multiplication. In a mild
generalisation of the original presentation of QTT, we also allow
sub-usaging via the \emph{reverse} ordering on the naturals. That is,
if a variable is marked as usage $n$ and $m \geq n$, then we can also
regard it as usage $m$. This makes the system more like affine linear
logic, since $m \sqsubseteq 0$\footnote{Reverse ordering!} for all $m$,
matching the system in \autoref{sec:linear-intro}. We do \emph{not}
have an unrestricted usage $\omega$, since this would allow the
possibility of unrestricted duplication, and hence violate our
polytime soundness properties.

\subsubsection{Contexts, Variables, and Conversion}

As we saw above, contexts in QTT are comprised of variable
$\stackrel\rho:$ type triples, where $\rho$ is a natural number
indicating how many times the variable $x$ is available for use in a
$\sigma = 1$ term. There are two operations on raw contexts: scaling
$\pi\Gamma$, which multiplies each $\rho$ in $\Gamma$ by $\pi$, and
addition $\Gamma_1 + \Gamma_2$, which adds two contexts' usage
annotations assuming that the lengths and types are equal. Contexts
are ordered pointwise $\Gamma' \sqsubseteq \Gamma$ on the usage
annotations (which is the \emph{reverse} ordering on naturals) The
basic usage-annotation discipline of QTT is demonstrated by the
context formation and variable rules:
\begin{mathpar}
  \inferrule*
  { }
  {\epsilon~\isctxt}

  \inferrule*
  {\Gamma~\isctxt \\ 0\Gamma \vdash S~\istype}
  {\Gamma, x \stackrel\rho: S~\isctxt}

  \inferrule*
  {0\Gamma, x \stackrel\sigma: S, 0\Gamma'~\isctxt}
  {0\Gamma, x \stackrel\sigma: S, 0\Gamma' \vdash x \stackrel\sigma: S}

  \inferrule*
  {\Gamma \vdash M \stackrel\sigma: S \\ \Gamma' \sqsubseteq \Gamma}
  {\Gamma' \vdash M \stackrel\sigma: S}
\end{mathpar}
As with most dependent type theories, contexts are built inductively
from the empty context $\epsilon$ and extension of a context by a
variable with a type that is well-formed in the preceding
context. Usage annotations $\rho$ on variables are arbitrary, and
types are always judged in a $0$-annotated context. The variable rule
marks unused variables as usage $0$ and the selected variable with
usage $\sigma$.

As usual, definitional equality of types impacts typing of terms via
the conversion rule:
\begin{displaymath}
  \inferrule*
  {\Gamma \vdash M : S \\ 0\Gamma \vdash S \equiv T~\istype}
  {\Gamma \vdash M : T}
\end{displaymath}
Like type formation, definitional equality of types always takes place
in $0$-d contexts. We will describe the definitional equality rules for
terms of each type as we introduce them. In QTT, it is possible for
the definitional equality of terms to differ between the $\sigma = 0$
and $\sigma = 1$ fragments, as we will see below.

\subsubsection{$\Pi$- and $\Sigma$-types}

QTT's $\Pi$-types have the form $(x \stackrel\rho: S) \to T$,
indicating functions that, in the $\sigma = 1$ fragment, use their
arguments $\rho$-many times. The formation, introduction and
elimination rules are similar to the standard ones, except for the
addition of usage annotations:
\begin{mathpar}
  \inferrule*
  {0\Gamma \vdash S~\istype \\ 0\Gamma, x \stackrel0: S \vdash T~\istype}
  {0\Gamma \vdash (x \stackrel\rho: S) \to T~\istype}

  \inferrule*
  {\Gamma, x \stackrel{\sigma\rho}: S \vdash M \stackrel\sigma: T}
  {\Gamma \vdash \lambda x.M \stackrel\sigma: (x \stackrel\rho: S) \to T}

  \inferrule*
  {\Gamma_1 \vdash M \stackrel\sigma: (x \stackrel\rho: S) \to T \\
    \Gamma_2 \vdash N \stackrel{\sigma'}: S \\
    0\Gamma_1 = 0\Gamma_2 \\
    \sigma' = 0 \Leftrightarrow (\rho = 0 \lor \sigma = 0)}
  {\Gamma_1 + \rho\Gamma_2 \vdash M\,N \stackrel\sigma: T[N/x]}
\end{mathpar}
The side conditions on the elimination rule state that (i) both
$\Gamma_1$ and $\Gamma_2$ erase to the same context, so their sum is
defined; and (ii) the argument $N$ is erased (i.e., $\sigma' = 0$) iff
either the function does not use its argument, or we are in the
$\sigma = 0$ fragment and everything is being erased. In the
following, when we write $S \to T$ for a non-dependent function type,
we mean that the argument is to be used linearly:
$(x \stackrel1: S) \to T$, where $x$ does not appear in
$T$. $\Pi$-types support the usual $\beta\eta$ definitional
equalities in both the $\sigma = 0$ and $\sigma = 1$ fragments.

$\Sigma$-types are a little more involved, and demonstrate the
flexibility in QTT in allowing additional power in the $\sigma = 0$
fragment where we do not need to care about polytime
realisability. Formation and introduction are given by the rules:
\begin{mathpar}
  \inferrule*
  {0\Gamma \vdash S~\istype \\ 0\Gamma, x \stackrel0: S \vdash T~\istype}
  {0\Gamma \vdash (x \stackrel\pi: S) \otimes T~\istype}

  \inferrule*
  {\Gamma_1 \vdash M \stackrel{\sigma'}: S \\
    \Gamma_2 \vdash N \stackrel\sigma: T[M/x] \\\\
    0\Gamma_1 = 0\Gamma_2 \\
    \sigma' = 0 \Leftrightarrow (\pi = 0 \lor \sigma = 0)}
  {\pi\Gamma_1 + \Gamma_2 \vdash (M, N) \stackrel\sigma: (x \stackrel\pi: S) \otimes T}
\end{mathpar}
As with $\Pi$-types, the first component of a $\Sigma$-type is
annotated with a usage for how many times it can be used, and this is
respected by the introduction rule. Elimination of $\Sigma$-types
depends on whether we are in the $\sigma = 0$ fragment or not. In the
$\sigma = 0$ fragment, we are free to disregard usage restrictions,
and use projections as normal:
\begin{mathpar}
  \inferrule*
  {\Gamma \vdash M \stackrel0: (x \stackrel\pi: S) \otimes T}
  {\Gamma \vdash \mathrm{fst}(M) \stackrel0: S}

  \inferrule*
  {\Gamma \vdash M \stackrel0: (x \stackrel\pi: S) \otimes T}
  {\Gamma \vdash \mathrm{snd}(M) \stackrel0: T[\mathrm{fst}(M)/x]}
\end{mathpar}
$\Sigma$-types are unrestricted in the $\sigma = 0$ fragment, and we
can use them as normal for type-theoretic constructions.  In the
$\sigma = 1$ fragment, we must take into account the resource content
of objects and use a pattern matching construct; the dependently typed
analogue of the $\otimes$-eliminator in \autoref{sec:linear-intro}:
\begin{mathpar}
  \inferrule*
  {0\Gamma, z \stackrel0: (x \stackrel\pi: A) \otimes B \vdash C \\
    \Gamma_1 \vdash M \stackrel\sigma: (x \stackrel\pi: A) \otimes B \\
    \Gamma_2, x \stackrel{\sigma\pi}: A, y \stackrel{\sigma}: B \vdash N \stackrel\sigma: C[(x,y)/z]\\
    0\Gamma_1 = 0\Gamma_2}
  {\Gamma_1 + \Gamma_2 \vdash \mathrm{let}~(x,y) = M~\mathrm{in}~N \stackrel\sigma: C[M/z]}
\end{mathpar}
QTT also supports a unit type $I$ with constructor $*$ and
pattern-matching \cite{atkey18qtt}. $\Sigma$- and $I$ types support
the usual $\beta\eta$ definitional equalities in the $\sigma = 0$
fragment (e.g., $\mathrm{fst}(M,N) \equiv M$), but only $\beta$
equalities (i.e.,
$\mathrm{let}~(x,y) = (M_1,M_2)~\mathrm{in}~N \equiv N[M_1/x,M_2/y]$)
in the $\sigma = 1$ fragment. It would also be sound to support
commuting conversions \cite{Barber1996} for $\mathrm{let}$ in the
$\sigma = 1$ fragment, but this would likely bring complications for
implementation.

\subsubsection{The Identity Type}

QTT also supports an extensional equality type with equality reflection:
\begin{mathpar}
  \inferrule*
  {0\Gamma \vdash S~\istype \\
    0\Gamma \vdash M \stackrel0: S \\
    0\Gamma \vdash N \stackrel0: S}
  {0\Gamma \vdash M =_S N~\istype}

  \inferrule*
  {\Gamma \vdash M \stackrel\sigma: S}
  {\Gamma \vdash \conRefl(M) \stackrel\sigma: M =_S M}

  \inferrule*
  {\Gamma \vdash N \stackrel0: M_1 =_S M_2}
  {\Gamma \vdash M_1 \equiv M_2 \stackrel0: S}
\end{mathpar}
The equality type also has an $\eta$ rule demonstrating $\conRefl(M)$
as the only proof of equality \cite{hofmann97syntax}. Note that
equality reflection only targets the $\sigma = 0$ fragment, we cannot
use equality reflection to convey realisability information.

\subsubsection{Universe} QTT has universe types $\mathsf{U}$, as in
standard type theory \cite{atkey18qtt}. For our examples below, we do
not explicitly mark the use of terms of type $\mathsf{U}$ as types --
i.e., we use a Russell-style presentation. Universes are where the
definitional equality on terms affects the definitional equality on
types.

\subsubsection{Data Types} QTT, as we have presented it so far, has no
interesting base types to perform computation on. Following our
presentation of the simply typed linear systems in
\autoref{sec:linear-intro}, we add two kinds of datatype to
QTT. First, in \autoref{sec:noniter-qtt}, we add non-iterable
datatypes that allow construction and case analysis, but no
recursion. Then, in \autoref{sec:cons-free-qtt} we describe how to
extend QTT to be a dependently typed adaptation of the \ConsFree{}
system of \autoref{sec:cons-free-intro} by adding a type of iterable
naturals. In \autoref{sec:lfpl-qtt} we apply the same treatment to
the LFPL-style system.

\subsection{Non-iterable Data Types}
\label{sec:noniter-qtt}

\subsubsection{Booleans} The boolean type for QTT was described in
\cite{atkey18qtt}. Booleans offer no possibility for iteration, but it
is useful to see how the QTT rules extend the simply typed rules from
\autoref{sec:linear-case-analysis} before moving to more complex
types.
\begin{mathpar}
  \inferrule*
  {\Gamma~\isctxt}
  {\Gamma \vdash \BoolTy~\istype}

  \inferrule*
  {\Gamma~\isctxt}
  {0\Gamma \vdash \cTrue, \cFalse \stackrel\sigma: \BoolTy}

  \inferrule*
  {0\Gamma_1, x \stackrel0: \BoolTy \vdash P~\istype \\
    \Gamma_1 \vdash M \stackrel\sigma: \BoolTy \\
    \Gamma_2 \vdash N_t \stackrel\sigma: P[\cTrue/x] \\
    \Gamma_2 \vdash N_f \stackrel\sigma: P[\cFalse/x] \\
    0\Gamma_1 = 0\Gamma_2}
  {\Gamma_1 + \Gamma_2 \vdash \If_{x.P}\: M \: \Then \: N_t \: \Else \: N_f \stackrel\sigma: P[M/x]}
\end{mathpar}
The introduction rules for booleans both use a $0$-d context,
indicating that construction of boolean values is free. Elimination of
booleans via a dependently typed if-then-else is more subtle with its
resource usage. The boolean to be eliminated must be constructed in a
context $\Gamma_1$, while the two branches are constructed in context
$\Gamma_2$. Since only one of the branches will be used, sharing
resources between the branches is expected. Booleans and their
eliminator obey the usual $\beta$ laws for definitional equality:
$\If_{x.P}\: \cTrue \: \Then \: N_t \: \Else \: N_f \equiv N_t$, and
similarly for $\cFalse$.

One might wonder how, since constructing booleans is $0$-cost by their
introduction rules, the $\Gamma_1$ context will ever be non-$0$. This
is resolved by observing that booleans may be the output of processes
that consume time (e.g., the iteration constructs defined below), and
so $\Gamma_1$ will represent a requirement that the necessary resource
is provided.

\subsubsection{Lists} Lists are a little more complex than booleans,
because the $\cCons$ constructor takes two arguments, so their
resource usage must be combined. The type formation and introduction
rules are as follows:
\begin{mathpar}
  \inferrule*
  {0\Gamma \vdash T~\istype}
  {0\Gamma \vdash \ListTy(T)~\istype}

  \inferrule*
  {\Gamma \vdash T~\istype}
  {0\Gamma \vdash \cNil \stackrel\sigma: \ListTy(T)}

  \inferrule*
  {\Gamma_1 \vdash M \stackrel\sigma: T \\
    \Gamma_2 \vdash N \stackrel\sigma: \ListTy(T) \\
    0\Gamma_1 = 0\Gamma_2}
  {\Gamma_1 + \Gamma_2 \vdash \cCons(M,N) \stackrel\sigma: \ListTy(T)}
\end{mathpar}
Lists do have the potential for iteration by their recursive nature,
but in order to ensure the polytime complexity guarantees we only permit
matching without recursion in the $\sigma = 1$ fragment. Here is the rule for
dependently typed case analysis on lists, which also obeys the usual
$\beta$-equalities for case analysis, analogous to the ones for
booleans:
\begin{mathpar}
  \inferrule*
  {0\Gamma_1, x \stackrel0: \ListTy(T) \vdash P~\istype \\
    \Gamma_1 \vdash M \stackrel\sigma: \ListTy(T) \\
    \Gamma_2 \vdash N_1 \stackrel\sigma: P[\cNil/x] \\
    \Gamma_2, h \stackrel\sigma: T, t \stackrel\sigma: \ListTy(T) \vdash N_2 \stackrel\sigma: P[\cCons(h,t)/x] \\
    0\Gamma_1 = 0\Gamma_2}
  {\Gamma_1 + \Gamma_2 \vdash \Match_{x.P}\,M\,\{\,\cNil \mapsto N_1; \cCons(h,t) \mapsto N_2\,\} \stackrel\sigma: P[M/x]}
\end{mathpar}
In the $\sigma = 0$ fragment, however, we are free to iterate on lists
because computations in this fragment are only meant for type-level
computation, not for the program itself. Put another way, the type
checker may perform arbitary recursion on lists to type check the
program, but the program itself may not do so without correctly
accounting its costs as described in the following sections. The
$\sigma = 0$ fragment recursor for lists has the following typing
rule, which is the standard dependent eliminator for lists except that
everything annotated as $0$ usage.
\begin{mathpar}
  \inferrule*
  {0\Gamma, x \stackrel0: \ListTy(T) \vdash P~\istype \\
    0\Gamma \vdash M \stackrel0: \ListTy(T) \\
    0\Gamma \vdash N_1 \stackrel0: P[\cNil/x] \\
    0\Gamma, h \stackrel0: T, t \stackrel0: \ListTy(T), p \stackrel0: P[t/x] \vdash N_2 \stackrel0: P[\cCons(h,t)/x]}
  {0\Gamma \vdash \mathrm{recList}_{x.P}\,M\,\{\,\cNil \mapsto N_1; \cCons(h,t;p) \mapsto N_2\,\} \stackrel0: P[M/x]}
\end{mathpar}
This eliminator also obeys the usual $\beta$-equality laws for a list
eliminator, using the resource freedom of the $\sigma = 0$ fragment to
duplicate the $N_2$ term in the $\cCons$ case.

\subsection{Cons-free Natural Numbers and their Recursor}
\label{sec:cons-free-qtt}

The datatypes of the previous section still only allow us to write
programs in the $\sigma = 1$ fragment that are constant time in the
size of their input. As with the simply typed linear system, if we are
handed a list of an unknown length, we can only explore it to a fixed
depth, determined statically by the program. To write programs that do
work proportional to the size of their input, we need some form of
iterable datatype. In both our \ConsFree{} and LFPL-style QTT systems,
we use a natural number datatype.

The \ConsFree{} system cannot allow the programmer to construct natural
numbers in the $\sigma = 1$ fragment, as this would violate the
complexity guarantees. However, we can use the flexibility of QTT to
allow free construction of naturals in the $\sigma = 0$ fragment,
which allows us to use natural numbers freely in types. Therefore, we
have the following introduction rules, only usable in the $\sigma = 0$
fragment:
\begin{mathpar}
  \inferrule*
  {\Gamma~\isctxt}
  {\Gamma \vdash \conZero \stackrel0: \tyNat}

  \inferrule*
  {\Gamma \vdash M \stackrel0: \tyNat}
  {\Gamma \vdash \conSucc(M) \stackrel0: \tyNat}
\end{mathpar}
The cons-free system allows free duplication of complete natural
numbers. This is accomplished by a special construct copying the
simply linear typed rule we gave above:
\begin{mathpar}
  \inferrule*
  {\Gamma \vdash M \stackrel\sigma: \tyNat}
  {\Gamma \vdash \mathrm{dupNat}(M) \stackrel\sigma: \tyNat \otimes \tyNat}
\end{mathpar}
Anyone who has reasoned about the metatheory of, or implemented a type
checker for, dependent types will view this rule with unease as it
appears to grant the ability to construct non-canonical values of pair
type, and consequently generate non-canonical naturals. We fix this by
adding an equational rule to the $\sigma = 0$ fragment, ensuring
definitionally that $\mathrm{dupNat}$ acts as its name implies:
\begin{mathpar}
  \inferrule*
  {\Gamma \vdash M \stackrel0: \tyNat}
  {\Gamma \vdash \mathrm{dupNat}(M) \equiv (M,M) \stackrel0: \tyNat \otimes \tyNat}
\end{mathpar}
Note that this rule is well-typed by the $0$-needs-$0$ property of
QTT, and the fact that $0+0=0$.

The eliminator for these natural numbers takes the following
form. Disregarding the usage annotations, it has the same type
structure as the normal dependently typed recursor for naturals:
\begin{displaymath}
  \mprset{flushleft}
  \inferrule*
  {0\Gamma, x \stackrel0: \tyNat \vdash P~\istype \\\\
    \Gamma \vdash M \stackrel\sigma: \tyNat \\\\
    0\Gamma \vdash N_z \stackrel\sigma: P[\conZero/x] \\\\
    0\Gamma, n \stackrel0: \tyNat, p \stackrel\sigma: P[n/x] \vdash N_s \stackrel\sigma: P[\conSucc(n)/x]}
  {\Gamma \vdash \tmRec_{x.P}\,M\,\{\conZero \mapsto N_z; \conSucc(n;p) \mapsto N_s\} \stackrel\sigma: P[M/x]}
\end{displaymath}
In the successor case, $N_s$, there are two bound variables: $n$ for
the natural number and $p$ for its induction hypothesis. Note that $n$
is required to be usage $0$ no matter what $\sigma$ is. We need the
variable $n$ to be present in order to correctly type the induction
hypothesis and the conclusion, but it must be marked as usage $0$ to
ensure that the resources captured by the number are not duplicated.

This eliminator cannot have a $\beta$-equality in the $\sigma = 1$
fragment because there is no way to construct any natural numbers to
iterate on in this fragment. In the $\sigma = 0$ fragment, this
eliminator obeys the usual $\beta$-equality laws for a natural number
recursor. This allows us to use it to compute and reason about
operations on naturals in this fragment.

The reader is invited to compare this dependently typed rule with the
simply typed linear version in \autoref{sec:cons-free-intro}. Removing
the $0$-annotated parts of the rule, and the type dependency, yield
the exact same rule. Conversely, when $\sigma = 0$, this rule is
identical (up to $0$-annotations) to the usual dependently typed
recursor for natural numbers, and so we can use it in the types to
prove properties of programs just as we do in normal type theory. We
will see in \autoref{sec:consfree-sound} that this rule is realisable
by polynomial-time computation, and so is sound for polynomial time.

\subsection{LFPL-style Diamonds, Natural Numbers, and a Recursor that Gives Back}
\label{sec:lfpl-qtt}

As explained in \autoref{sec:lfpl-intro}, the LFPL system differs from
the \ConsFree{} system in that it is possible to construct natural
numbers (and other iterable datatypes), provided one has the necessary
diamonds to pay for the construction. As with the natural number type
in the \ConsFree{} system, it ought not be possible to construct
diamonds in the $\sigma = 1$ fragment, as this would amount to the
free distribution of diamonds to all which would lead to a collapse in
the complexity guarantees of the system. It is possible construct
diamonds in the $\sigma = 0$, though:
\begin{mathpar}
  \inferrule*
  {\Gamma~\isctxt}
  {0\Gamma \vdash \Diamond~\istype}

  \inferrule*
  {\Gamma~\isctxt}
  {0\Gamma \vdash * \stackrel0: \Diamond}

  \inferrule*
  {\Gamma \vdash M \stackrel0: \Diamond}
  {\Gamma \vdash M \equiv * \stackrel0: \Diamond}
\end{mathpar}
The $\Diamond$ type also supports an $\eta$-rule in the $\sigma = 0$
fragment, indicating that, in this fragment, it acts the same as a
unit type. This allows us to freely use diamonds in types, and to
not have to care about the identity of particular diamonds, since by
this rule all diamonds are definitionally
equal\footnote{\emph{Fungible}, if one wishes to use a monetary
  metaphor.}.

Construction of natural numbers now requires a $\Diamond$ for
$\conZero$ and a $\Diamond$ and a predecessor for $\conSucc$:
\begin{mathpar}
  \inferrule*
  {\Gamma \vdash M \stackrel\sigma: \Diamond}
  {\Gamma \vdash \conZero(M) \stackrel\sigma: \tyNat}

  \inferrule*
  {\Gamma_1 \vdash M \stackrel\sigma: \Diamond \\
    \Gamma_2 \vdash N \stackrel\sigma: \tyNat \\
    0\Gamma_1 = 0\Gamma_2}
  {\Gamma_1 + \Gamma_2 \vdash \conSucc(M,N) \stackrel\sigma: \tyNat}
\end{mathpar}
In the $\sigma = 0$ fragment, we can construct $\Diamond$s for free,
and so construct natural numbers freely as well just as we did for the
\ConsFree{} system above.

The dependently typed recursor for LFPL-style natural numbers again
augments the simply typed linear recursor from
\autoref{sec:lfpl-intro} with dependency information:
\begin{mathpar}
  \mprset{flushleft}
  \inferrule*
  {0\Gamma, x \stackrel0: \tyNat \vdash P~\istype \\\\
    \Gamma \vdash M \stackrel\sigma: \tyNat \\\\
    0\Gamma, d \stackrel\sigma: \Diamond \vdash N_z \stackrel\sigma: P[\conZero(*)/x] \\\\
    0\Gamma, d \stackrel\sigma: \Diamond, n \stackrel0: \tyNat, p \stackrel\sigma: P[n/x] \vdash N_s \stackrel\sigma : P[\conSucc(*,n)/x]}
  {\Gamma \vdash \tmRec\,M\,\{\conZero(d) \mapsto N_z; \conSucc(d,n;p) \mapsto N_s\} \stackrel\sigma: P[M/x]}
\end{mathpar}
We have used $* : \Diamond$ as the value in the types for the zero and
successor cases. By the $\eta$-rule for diamonds, we could have
equally well used the $d$ variable that is in scope in each case.

Unlike the natural number iterator in the \ConsFree{} system, this
iterator has $\beta$-equalities in both the $\sigma = 0$ and
$\sigma = 1$ fragments. In the $\conSucc$ case, for example, we have:
\begin{displaymath}
  \begin{array}{cl}
    & \tmRec\,(\conSucc(M_d,M_n))\,\{\conZero(d) \mapsto N_z; \conSucc(d,n;p) \mapsto N_s\} \\
    \equiv & N_s[M_d/d, M_n/n, \tmRec\,M_n\,\{\conZero(d) \mapsto N_z; \conSucc(d,n;p) \mapsto N_s\}/p]
  \end{array}
\end{displaymath}
Note that the fact that the variable $n$ in the $N_s$ term is
annotated $0$, which allows us to use $M_n$ twice even when we are in
the $\sigma = 1$ fragment.

Just as for the \ConsFree{} system iterator above, in the $\sigma = 0$
fragment this rule is identical to the usual dependently typed
recursor for the natural numbers, so it can be used in the types to
reason about programs. Moreover, we will see in
\autoref{sec:lfpl-sound} that this rule is also sound for polynomial
time in a system with $\Diamond$s.

\subsection{Reflection of Realisability}
\label{sec:qtt-reflection}

Our final addition to QTT is \emph{reflection of realisability}. In
QTT thus far, it has been possible to reason about the non-resourced
behaviour of programs. This is because the $0$-ing process moving from
the $\sigma = 1$ fragment to the $\sigma = 0$ fragment erases all
resource information. This is sufficient for reasoning about the
extensional behaviour of programs via types, but it is useful to be
able to make statements like ``this function is realisable in
polynomial time'' in the types of QTT, something that is not currently
possible with the system we have seen so far.

We remedy this by adding a \emph{realisable} type to QTT with the
following type formation and introduction and elimination rules:
\begin{mathpar}
  \inferrule*
  {0\Gamma \vdash A~\istype}
  {0\Gamma \vdash \Rtype(A)~\istype}

  \inferrule*
  {0\Gamma \vdash M \stackrel1: A}
  {0\Gamma \vdash \rIntro(M) \stackrel\sigma: \Rtype(A)}

  \inferrule*
  {\Gamma \vdash M \stackrel\sigma: \Rtype(A)}
  {\Gamma \vdash \rElim(M) \stackrel{\sigma'}: A}
\end{mathpar}
Intuitively, the type $\Rtype(A)$ is inhabited whenever the type $A$
is realisable in the $\sigma = 1$ fragment of the system. In
particular, the type $\Rtype(\tyNat \to \tyNat)$ is the type of all
realisable functions from natural numbers to natural numbers. In the
polynomial time systems we are concerned with here, this is exactly
the type of polynomial functions. Note that in the introduction rule,
the premise is required to be in the $\sigma = 1$ fragment, to ensure
that the type is realisable, while in the elimination rule, the
conclusion is in an arbitrary fragment $\sigma'$. This flexibility is
require to maintain the admissibility of the $0$-ing rule.

The equality rules for $\Rtype$ state that the two operations are
mutually inverse: $\rIntro(\rElim(M)) \equiv M$, in both fragments,
and $\rElim(\rIntro(M)) \equiv M$ in the $\sigma = 0$ fragment. By
congruence, the $\sigma = 1$ fragment's definitional equality affects
the definitional equality of the $\sigma = 0$ fragment via the
$\rIntro(-)$ constructor.

With just the rules given here, the type $\Rtype(A)$ is no more than a
statement that a given type is realisable with a polytime
implementation. This is enough to do the constructions that we present
in the next section, e.g., that polytime functions are closed under
composition, but one could imagine stronger reflection principles that
allow deeper logical consequences of polytime realisability to be
proved internally. We discuss this further in
\autoref{sec:towards-synthetic-complexity-theory}.

Readers familiar with \citet{Benton94}'s Linear/Non-linear system will
note that the $\Rtype(A)$ constructor is the QTT analogue of the right
adjont $G$ type constructor in that system. The $\Sigma$-types play
the role of the left adjoint $F$ types, in a similar way to the
dependent linear type system of \citet{KrishnaswamiPB15}.

\section{Programming and Proving with Polytime}
\label{sec:programming-polytime}

We now explore the possibilities afforded by the combination of
polytime guarantees with the specification expressivity of dependent
types.

\subsection{Building Data Types}
\label{sec:prog-datatypes}

We have only defined an iterable natural number datatype for both of
our systems above. We could extend both systems to include further
iterable inductive types, although in the \ConsFree{} system this
is not particularly useful due to the prohibition of
construction. However, sticking with just the natural numbers, we can
use the power of dependent types with a universe to create further
datatypes whose size is measured by some iterable natural
number. Iteration on the size yields iteration over the full
datastructure. For example, in the LFPL system, we can define a type
of iterable lists by pairing a size with a type of elements defined by
recursion on the size:
\begin{displaymath}
  \mathrm{IList}\,A = (n \stackrel1: \tyNat) \otimes \left(\mathrm{rec}_{x.\mathsf{U}}\,n\,\{ \conZero(d) \mapsto I; \conSucc(d,n;p) \mapsto A \otimes p \}\right)
\end{displaymath}
The $\cNil$ and $\cCons$ constructors can now be defined in terms of
$\conZero$ and $\conSucc$, provided the caller supplies sufficient
$\Diamond$s. These definitions live in the $\sigma = 1$ fragment, so
we annotate them appropriately:
\begin{mathpar}
  \begin{array}{l}
    \mathrm{nil} \stackrel1: \Diamond \to \mathrm{IList}\,A \\
    \mathrm{nil}\,d = (\conZero(d), *)
  \end{array}

  \begin{array}{l}
    \mathrm{cons} \stackrel1: \Diamond \to A \to \mathrm{IList}\,A \to \mathrm{IList}\,A \\
    \mathrm{cons}\,d\,x\,\mathit{xs} = \mathrm{let}\,(n,\mathit{elems}) = \mathit{xs}\,\textrm{in}\,(\conSucc(d,n),(x,\mathit{elems}))
  \end{array}
\end{mathpar}
Using the LFPL iterator it is also possible to construct a dependently
typed iterator for $\mathrm{IList}\,A$ values. Unfortunately, the
current types of the LFPL system are not sufficient to type this as a
function, as we have no way of stating that the successor case must be
arbitrarily duplicable. Lifting this restriction by means of some
modality is future work. The typing rule for the derived list iterator
is:
\begin{mathpar}
  \mprset{flushleft}
  \inferrule*
  {0\Gamma \vdash A~\istype \\
    0\Gamma, x \stackrel0: \mathrm{IList}\,A \vdash P~\istype \\\\
    \Gamma \vdash M \stackrel\sigma: \mathrm{IList}\,A \\\\
    0\Gamma, d \stackrel1: \Diamond \vdash N_1 \stackrel\sigma: P[\mathrm{nil}(*)/x] \\\\
    0\Gamma, d \stackrel1: \Diamond, x \stackrel\sigma: A, \mathrm{xs} \stackrel0: \mathrm{IList}\,A, p \stackrel\sigma: P[\mathit{xs}/x] \vdash N_2 \stackrel\sigma : P[\mathrm{cons}(*,x,\mathit{xs})/x]}
  {\Gamma \vdash \tmRec_{x.P}\,M\,\{\mathrm{nil}(d) \mapsto N_1; \mathrm{cons}(d,x,\mathit{xs};p) \mapsto N_2\} \stackrel\sigma: P[M/x]}
\end{mathpar}
Note that, in the cons case, we have access to the result of iterating
over the tail of the list ($p$), but not to the actual tail of the
list ($\mathit{xs}$).

With our list iterator, it is now possible to write interesting
polytime programs. For example, the example used by \citet{Hofmann03}
to demonstrate the expressivity of LFPL is insertion sort. First we
define insertion of a natural into a sorted list:
\begin{displaymath}
  \mathrm{insert} \stackrel1: \Diamond \to \tyNat \to \mathrm{IList}\,\tyNat \to \mathrm{IList}\,\tyNat
\end{displaymath}
which requires some ingenuity to write to handle the case where we
find the place to insert the item and need access to the remainder of
the list. Note that, also, the function consumes a $\Diamond$ to
construct the new element of the output list, and also that the items
in the list are themselves iterable natural numbers. This is needed to
account for the comparisons between elements.

Insertion sort is repeated insertion of elements from an original list
into a new list. The new list is constructed from the $\Diamond$s
yielded by the original list:
\begin{displaymath}
  \mathrm{insertionSort} \stackrel1: \mathrm{IList}\,A \to \mathrm{IList}\,A
\end{displaymath}
The immediate benefit of dependent types in this situation is that it
is now possible to state and prove the correctness property of this
sorting procedure. Using the fact that the $\sigma = 0$ fragment of
QTT is exactly normal type theory, we can use normal dependently typed
programmming techniques to establish:
\begin{displaymath}
  \mathrm{insertionSortCorrect} \stackrel0: (\mathit{xs} \stackrel1: \mathrm{IList}\,A) \to \mathrm{Sorted}(\mathit{xs},\mathrm{insertionSort}\,\mathrm{xs})
\end{displaymath}
where $\mathrm{Sorted}(x,y)$ is some predicate stating that $y$ is a
sorted permutation of $x$.  Note that, despite the $1$ annotation on
the $\Pi$-type here, we are free to duplicate $\mathit{xs}$ because
types are constructed in the $\sigma = 0$ fragment.

\subsection{Polytime Problems}
\label{sec:polytime-problems}

Define a decision problem to be a pair $(A, P)$, where $A$ is a type
in the universe $\mathsf{U}$, and $P : A \to \mathsf{U}$ is a
predicate on $A$. For what follows, we are only interested in whether
or not $P\,a$ is inhabited for each $a$. Therefore, we use
$P \Leftrightarrow Q$ to stand for equi-inhabitation of two
$P, Q : \mathsf{U}$, i.e.,
$P \Leftrightarrow Q \equiv (P \to Q) \times (Q \to P)$.

We can use the reflection type former defined in
\autoref{sec:qtt-reflection} to define a predicate on decision
problems that establishes whether or not they are polytime decision
problems. Specifically, we can state that there is a polytime
realisable boolean-value predicate that reports true exactly when the
given element of $a$ is in the predicate:
\begin{displaymath}
  \mathrm{PTIME}(A,P) =
    (f \stackrel1: \mathbf{R}(A \to \BoolTy)) \otimes \left((a \stackrel1: A) \to (\mathbf{R}^{-1}(f)\, a = \cTrue) \Leftrightarrow P\,a\right)
\end{displaymath}
Thus, $\mathrm{PTIME}(A,P)$ is a logical proposition stating that the
decision problem $(A,P)$ is decidable in polytime. We make three notes
about this definition: (i) \emph{proofs} of $\mathrm{PTIME}(A,P)$, are
carried out in the $\sigma = 0$ fragment, where we have the full power
of Type Theory to aid us; (ii) this definition is intrinsic, in the
sense that, whichever of the polytime systems is chosen, proving that
a decision problem is solvable in polytime is a matter of programming,
without having to reason directly about machine models and step
counting; and (iii) we have defined problems to have arbitrary types
$A$ as domains, rather than bitstrings, and so the notion of size
attached to an input is intrinsic to the type $A$ chosen.

We can also declare a type of polytime \emph{reductions} between
problems. A problem $(A, P)$ can be polytime reduced to a problem
$(B,Q)$ if there is an inhabitant of the following type:
\begin{displaymath}
  (A,P) \stackrel{\textsc{Poly}}\Rightarrow (B,Q) = (f \stackrel1: \mathbf{R}(A \to B)) \otimes \left((a \stackrel1: A) \to Q(\mathbf{R}^{-1}(f)\,a) \Leftrightarrow P\,a\right)
\end{displaymath}
In words, there must be a polytime function $f$ that preserves and
reflects decisions. With this definition, it is possible to prove in
our systems that polytime computations are closed under polytime
reductions. We note that this definition is, up to the reflection
modality, the same as the definition of cartesian container morphism,
well known in dependent type theory \cite{AbbottAG05}, and speaks to a
general conception of containers as ``problem/solution'' pairings and
container morphisms as problem reductions.

\subsection{Polytime-based Complexity Classes}
\label{sec:classes}

The fact that we can characterise polytime decision problems is
perhaps to be expected from a system designed to capture polynomial
time realisable programs. However, we can go further to capture the
complexity classes of Non-deterministic Polynomial time (NP) and
Probabilistic Polynomial time (PP), both of which are based on
polytime. We do this by augmenting our polytime functions with
additional power in the form of computational effects.

\subsubsection{Non-deterministic Polynomial Time}
\label{sec:np-class}

To capture the complexity class NP, we use polynomial time programs
augmented with non-determinism, as one might expect. We will not need
to reason about equality of these non-deterministic programs, so we
can represent non-deterministic choices as binary trees. We suppose a
non-iterable datatype defined like so:
\begin{displaymath}
  \begin{array}{l}
    \textbf{data}\,\mathrm{ND}\,(A : \mathsf{U}) : \mathsf{U}\,\textbf{where} \\
    \quad \mathsf{return} : A \to \mathrm{ND}\,A \\
    \quad \mathsf{choice} : (\BoolTy \to \mathrm{ND}\,A) \to \mathrm{ND}\,A
  \end{array}
\end{displaymath}
The crucial point here is that the subtrees are represented as a
function $\BoolTy \to \mathrm{ND}\,A$. By the typing rules of QTT,
this means that the two branches of this function can share resources
(see the encoding of the additive product types by
\citet{atkey18qtt}). Thus, each branch of this tree can be explored in
polynomial time, but not the whole tree itself.

The type $\mathrm{ND}$ supports a monad interface via the usual free
monad construction, as well as an effect
$\mathrm{flip} \stackrel1: \mathrm{ND}\,\BoolTy$ providing access to a
bit of non-deterministic information. Thus a program of type
$A \to \mathrm{ND}\,B$ in the $\sigma = 1$ fragment will be a polytime
program with access to an oracle. In the $\sigma = 0$ fragment, we can
write a function that resolves non-determinism using a list of
booleans. This function returns $\mathsf{nothing}$ if the list of
booleans is insufficient to resolve all the $\mathsf{choice}$s:
\begin{displaymath}
  \mathrm{runWithOracle} \stackrel0: \mathrm{ND}\,A \to \ListTy(\BoolTy) \to \mathrm{Maybe}\,A
\end{displaymath}
With these definitions, we can define Non-deterministic Polynomial
time as a predicate on problems:
\begin{displaymath}
  \mathrm{NP}(A,P) =
  \begin{array}[t]{@{}l}
    (f \stackrel1: \mathbf{R}(A \to \mathrm{ND}(\BoolTy)))\, \otimes \\
    \, \left((a \stackrel1: A) \to \left((\mathit{bs} \stackrel1: \ListTy(\BoolTy)) \otimes (\mathrm{runWithOracle}\,(\mathbf{R}^{-1}(f)\,a)\,\mathit{bs} = \mathsf{just}\,\cTrue)\right) \Leftrightarrow P\,a\right)
  \end{array}
\end{displaymath}
Thus, a problem is in NP if there is a non-deterministic
boolean-valued polynomial time function that has a path to returning
$\cTrue$ exactly when the input satisfies the predicate. Moreover, it is a
quick matter of programming to see that problems in NP are closed
under the type of polytime reductions given above.

\subsubsection{Bounded-error Probabilistic Polynomial Time}

By changing the computation effects supplied to a program, we can
change the complexity class. To capture the class BPP of Bounded-error
Probabilistic Polynomial time \cite{AroraBarak09}, we use a
(non-iterable) data structure representing trees of probabilistic
choices, where $\mathbb{Q}[0,1]$ is some type of (non-iterable)
rationals in the closed interval $[0,1]$:
\begin{displaymath}
  \begin{array}{l}
    \textbf{data}\,\mathrm{Dist}\,(A : \mathsf{U}) : \mathsf{U}\,\textbf{where} \\
    \quad \mathsf{return} : A \to \mathrm{Dist}\,A \\
    \quad \mathsf{choice} : \mathbb{Q}[0,1] \to (\BoolTy \to \mathrm{Dist}\,A) \to \mathrm{Dist}\,A
  \end{array}
\end{displaymath}
As in the non-deterministic case, a function $A \to \mathrm{Dist}\,B$
in the $\sigma = 1$ fragment is a polytime probabilistic
computation. Again, the use of a function type here ensures that each
branch of the tree is constructable in polynomial time, not the whole
tree. In the $\sigma = 0$ fragment we can write a function that
computes the probability of a $\mathrm{Dist}\,\BoolTy$ computation
being true:
\begin{displaymath}
  \mathrm{probTrue} \stackrel0: \mathrm{Dist}\,\BoolTy \to \mathbb{Q}[0,1]
\end{displaymath}
We can now define the class of probabilistic polynomial time decision
problems, where the decider is allowed to make probabilistic choices
as long as it is correct with probability at least $\frac{2}{3}$:
\begin{displaymath}
  \mathrm{BPP}(A,P) =
  \begin{array}[t]{l}
    (f \stackrel1: \mathbf{R}(A \to \mathrm{Dist}(\BoolTy))) \, \otimes
    \left((a \stackrel1: A) \to (\mathrm{probTrue}\,(\mathbf{R}^{-1}(f)\,a) \geq \frac{2}{3}) \Leftrightarrow P\,a\right)
  \end{array}
\end{displaymath}
Again, problems in BPP are easily seen to be closed under polytime
reductions.

Probabilistic Polynomial time has previously been considered in the
setting of implicit computational complexity by
\citet{dallago_et_al:LIPIcs.MFCS.2021.35} and \citet{LagoT15}. In both
cases, they must build probabilistic choice into the language, and
have difficulty in directly capturing the class BPP due to its
semantic nature, where the correctness of implementation is
probabilistic. With a dependently-typed host language, adding
probabilistic choice as an effect and capturing the semantic
constraint of BPP is straightforward.

\section{Polytime Soundness via Realisability}
\label{sec:soundness}

\newcommand{\cstaccess}{1}
\newcommand{\cstmkclo}{1}
\newcommand{\cstapp}{1}
\newcommand{\cstmkpair}{1}
\newcommand{\cstmkunit}{1}
\newcommand{\cstTrue}{1}
\newcommand{\cstFalse}{1}
\newcommand{\cstLetpair}{1}
\newcommand{\cstSeq}{1}
\newcommand{\cstIf}{1}

\newcommand{\clo}[2]{\mathsf{clo}\langle #1 , #2 \rangle}
\newcommand{\synTrue}{\mathsf{true}}
\newcommand{\synFalse}{\mathsf{false}}

\newcommand{\ExpSet}{\mathcal{E}}
\newcommand{\ValSet}{\mathcal{V}}

\newcommand{\rplus}{\oplus}
\newcommand{\rzero}{\emptyset}

In this section and the next, we establish the polytime soundness of
our extensions of QTT by adapting a realisability method due to
\citet{dallago11realisability}. This approach is based on a three way
coupling between abstract mathematical elements (the \emph{what}),
values from a machine model (the \emph{how}), and resource potentials
(the \emph{fuel}). Each type in the system is defined as a three way
relation between these elements. The set of abstract elements depends
on the type being interpreted (e.g., types of natural numbers will be
defined in terms of the set $\mathbb{N}$). The machine model is fixed
across all types. We describe the particular machine model we use for
this paper in \autoref{sec:machine-model}. Potentials are arranged
into \emph{resource monoids} that we define in
\autoref{sec:resource-monoids}. Unlike \citet{dallago11realisability},
we explicitly construct realisers for inductive datatypes (both
iterable and non-iterable) instead of relying on second-order
polymorphic encodings and special $\oc$-style modalities. These
explicit constructions are essential to construct models of our systems.

\paragraph{Agda Formalisation} The key soundness results in this
section have been formalised in the Agda proof assistant
\cite{norell2008dependently}. The Agda formalisation can be found in
the associated artefact \cite{atkey2023polydep-artefact}. After each
definition and result we provide a pointer to the Agda modules where
the corresponding formalisation can be found, and note interesting
features of the mechanisation.

\subsection{Machine Model and Operational Semantics}
\label{sec:machine-model}

\begin{figure}
  \centering
  {\bf Syntax}
  \begin{displaymath}
    \begin{array}{lcl}
      i,j &\in&\mathbb{N} \\
      E \in \ExpSet &::=& \lambda E \mid * \mid (i, j) \mid \synTrue \mid \synFalse \mid i \mid \Let\,E_1\,\In\,E_2 \mid i \cdot j \mid \LetPair\,i\,\In\,E \mid \If\,i\,E_1\,E_2 \\
      V \in \ValSet &::=& \clo{E}{\eta} \mid * \mid (V_1, V_2) \mid \synTrue \mid \synFalse \\
      \eta &::= & [] \mid \eta :: V
    \end{array}
  \end{displaymath}

  \vspace{1em}

  {\bf Evaluation: Construction}
  \begin{mathpar}
    \inferrule* [right=MkClo]
    { }
    {\lambda E , \eta \Downarrow_{\cstmkclo} \clo{E}{\eta}}

    \inferrule* [right=MkUnit]
    { }
    {*, \eta \Downarrow_{\cstmkunit} *}

    \inferrule* [right=MkPair]
    {\eta[i] = V_1 \\ \eta[j] = V_2}
    {(i, j), \eta \Downarrow_{\cstmkpair} (V_1, V_2)}

    \inferrule* [right=MkTrue]
    { }
    {\synTrue, \eta \Downarrow_{\cstTrue} \synTrue}

    \inferrule* [right=MkFalse]
    { }
    {\synFalse, \eta \Downarrow_{\cstFalse} \synFalse}
  \end{mathpar}

  \vspace{1em}

  {\bf Evaluation: Variable access and Sequencing}
  \begin{mathpar}
    \inferrule* [right=Access]
    {\eta[i] = v}
    {i, \eta \Downarrow_{\cstaccess} v}

    \inferrule* [right=Seq]
    {E_1, \eta \Downarrow_{k_1} V \\
      E_2, (\eta :: V) \Downarrow_{k_2} V'}
    {\Let\,E_1\,\In\,E_2, \eta \Downarrow_{k_1 + \cstSeq + k_2} V'}
  \end{mathpar}

  \vspace{1em}

  {\bf Evaluation: Elimination}
  \begin{mathpar}
    \inferrule* [right=App]
    {\eta[i] = \clo{E}{\eta'} \\
      \eta[j] = V \\
      E , (\eta' :: \clo{E}{\eta'} :: V) \Downarrow_k V'}
    {(i \cdot j), \eta \Downarrow_{\cstapp + k} V'}

    \inferrule* [right=LetPair]
    {\eta[i] = (V_1, V_2) \\
      E, (\eta :: V_1 :: V_2) \Downarrow_k V}
    {\LetPair\,i\,\In\,E, \eta \Downarrow_{\cstLetpair + k} V}

    \inferrule* [right=IfTrue]
    {\eta[i] = \synTrue \\ E_1, \eta \Downarrow_k V}
    {\If\,i\,E_1\,E_2, \eta \Downarrow_{\cstIf + k} V}

    \inferrule* [right=IfFalse]
    {\eta[i] = \synFalse \\ E_2, \eta \Downarrow_k V}
    {\If\,i\,E_1\,E_2, \eta \Downarrow_{\cstIf + k} V}
\end{mathpar}
\caption{Language with CBV Big-step Costed Evaluation Semantics}
\Description{Syntax and evaluation rules of the untyped CBV
  $\lambda$-calculus used for realisers.}
  \label{fig:opsem}
\end{figure}

We demonstrate that every program that can be written in extensions of
QTT has the complexity bounds that we claim by translating QTT terms
into an untyped CBV $\lambda$-calculus with a costed operational
semantics. The syntax and rules of our target language are given in
\autoref{fig:opsem}.

Variables are represented as de Bruijn indicies $i, j$. Expressions
$E \in \ExpSet$ can be (anonymous) $\lambda$-abstractions, unit,
pairing and boolean values, variables, sequencing, application, pair
elimination, and conditionals. Note that, with the exception of
$\lambda$-abstraction and sequencing, expressions never contain nested
expressions; instead referring to variables already defined. Values
$V \in \ValSet$ can be closures $\clo{E}{\eta}$, where $\eta$ is an
environment for the closure, unit values, pairs and booleans.

Costed evaluation of expressions in environments is defined by a
big-step operational semantics $E, \eta \Downarrow_k V$, where $k$ is
the number of steps. For simplicity, all operations cost $1$ unit,
though this could be generalised to allow for different operations to
have different costs. We use $\eta[i]$ to access the $i$th variable in
the environment, counting from the right. The evaluation rules are
mostly as one would expect, except for the application rule which
includes a self reference to the closure being invoked, in order to
allow recursive definitions.

\paragraph{Agda Formalisation} The machine model is defined in the
Agda module \AgdaModule{MachineModel}. We use an intrinsically
well-scoped syntax, which ensures that all variable look up operations
are always well defined.

\subsection{Resource Monoids}
\label{sec:resource-monoids}

As we mentioned above, resource potentials are attached to values to
represent the amount of intrinsic potential they have to fuel
computation. Resource potentials are organised into resource
monoids. To be able to account for the combined potential attached to
composite data and programs (e.g., pairs, or functions applied to
arguments) we will require monoid structure on potentials. The action
of turning potential difference into fuel for computation will be
modelled by a difference function. Finally, we require that our
resource monoid contains sufficient elements to fuel constant time
operations. We gather these requirements into a formal definition as
follows, which is a slight reformulation of the resource monoids of
\citet{dallago11realisability}:

\begin{definition}
  A \emph{resource monoid} $M$ consists of:
  \begin{enumerate}
  \item A carrier set $|M|$, whose elements represent amounts of
    potential. We use Greek letters $\alpha$, $\beta$, $\gamma$ to
    denote elements of a resource monoid.
  \item Commutative monoid structure $(\rplus,\rzero)$ on $|M|$, so
    we can add potentials.
  \item a \emph{difference function} $M : |M| \times |M| \to \natinf$,
    where $\natinf$ is the natural numbers extended with a negative
    infinity $- \infty$ and $- \infty + k = -\infty$. A difference
    $M(\alpha, \beta) = k \in \mathbb{N}$ means that starting with
    potential $\alpha$ and ending with potential $\beta$ yields $k$
    units of fuel. A difference of $- \infty$ means that $\alpha$
    contains insufficient potential to reach $\beta$. Differencing
    must satisfy:
    \begin{enumerate}
    \item for all $\alpha$, $M(\alpha,\alpha) = 0$; and
    \item for all $\alpha, \beta, \gamma$,
      $M(\alpha, \beta) + M(\beta, \gamma) \leq M(\alpha, \gamma)$.
    \end{enumerate}
    The latter is a ``reverse triangle inequality'': the fuel
    recoverable by moving between potential levels $\alpha$ and
    $\gamma$ via $\beta$ may be less than the fuel recoverable
    moving from $\alpha$ to $\gamma$ directly.
  \item Differencing and the commutative monoid structure must satisfy:
    \begin{enumerate}
    \item $M(\alpha, \beta) \leq M(\alpha \rplus \gamma, \beta \rplus \gamma)$; and
    \item $M(\alpha, \rzero) = 0$.
    \end{enumerate}
  \item An \emph{accounting function}
    $\mathit{acct} : \mathbb{N} \to |M|$ such that for all $k$,
    $k \leq M(\mathit{acct}(k),\rzero)$.
  \end{enumerate}
\end{definition}

For any resource monoid $M$, we can define an action of $\mathbb{N}$
on $M$ as $n \cdot \alpha = \alpha \oplus \cdots \oplus \alpha$, where
the right-hand side has $n$ summands.

\paragraph{Alternative definition} Every resource monoid induces a
pre-ordering on its carrier set by $\alpha \leq \beta$ iff
$0 \leq M(\alpha, \beta)$. Taking this idea further, we can
reformulate a resource monoid as a symmetric monoidal category
enriched in the symmetric monoidal category $\natinf$, where the
monoid structure is addition. The conditions in the definition above
amount to the usual identity and composition laws for enriched
categories. With this reading, we can see the value $M(\alpha,\beta)$
when it is $\geq 0$ as the possibility of moving from $\alpha$ to
$\beta$ levels of potential resource with some amount of residual
resource emitted for computation; when it is $-\infty$, moving from
$\alpha$ to $\beta$ is not possible.

\paragraph{Agda Formalisation} Resource monoids are defined in the
module \AgdaModule{Algebra.ResourceMonoid}. We use a formulation
closer to the enriched category theory definition for the actual
formalisation, because it avoids having to treat equality in the
monoid structure separately from the induced preorder on
elements. Thinking of proofs involving the resource monoid as a
process of finding a composable sequence of morphisms in a category
was a helpful intuition when constructing the realisability model
below.

\subsubsection{Specific Resource Monoids}
\label{sec:specific-monoids}

The simplest example of a resource monoid is given by the natural
numbers $\mathbb{N}$, where each number stands directly an amount of
stored fuel.

\begin{definition}[Natural Number Resource Monoid]
  Monoid structure is given by normal addition. Differencing is
  defined as
  \begin{displaymath}
    \mathbb{N}(m,n) = \left\{
      \begin{array}{ll}
        m - n & m \geq n \\
        - \infty & \textrm{otherwise}
      \end{array}\right.
  \end{displaymath}
  and $\mathit{acct}(k) = k$. Note that this is the simplest possible
  resource monoid due to the requirement that the $\mathit{acct}$
  function must exist.
\end{definition}

\newcommand{\MaxPoly}{\mathrm{MaxPoly}}
\newcommand{\PlusPoly}{\mathrm{PlusPoly}}

The differencing operator of the natural number resource monoid can
only supply as much fuel as is contained in the potential.  For the
two polynomial time systems, we need more sophisticated structures,
both originally presented by Dal Lago and Hofmann. The fundamental
idea with both is to represent potentials as pairs $(m,p)$, where $m$
is a natural number and $p$ is a polynomial. The $m$ tracks the
``size'' of data as it pertains to the number of times an operation
will be repeated by iterating over it --- for example, an iterable
natural number will have size equal to itself, but a non-iterable
natural number may be assigned zero size. The polynomial $p$ tracks
the complexity of a program as a function of the size of the
input. This leads to a differencing operator that evaluates the
polynomial with the size of the data:

\begin{definition}[Polynomial Resource Monoids]\label{def:polynomial-resource-monoids}
  The \emph{Max-Polynomial} resource monoid $\MaxPoly$ has carrier set
  consisting of pairs $(m,p)$ where $m$ is natural number and $p$ is a
  polynomial with natural number coefficients. Addition of elements is
  defined as $(m,p) \oplus (n,q) = (m \sqcup n, p+q)$, where $\sqcup$
  is the max operator, with $\emptyset = (0,0)$. Difference is defined
  as:
  \begin{displaymath}
    \MaxPoly((m,p),(n,q)) = \left\{
      \begin{array}{ll}
        p(m) - q(m)&m \geq n \textrm{ and } \forall k \geq m. p(k) \geq q(k) \\
        - \infty & \textrm{otherwise}
      \end{array}
    \right.
  \end{displaymath}
  $\MaxPoly$ accounts for constant time with constant polynomials:
  $\mathit{acct}(k) = (0,\lambda x.k)$.

  The \emph{Plus-Polynomial} resource monoid $\PlusPoly$ is defined
  the same way as $\MaxPoly$ except that the monoid addition adds the
  natural number components instead of taking their maximum:
  $(m,p) \oplus (n,q) = (m + n, p+q)$.
\end{definition}

It is perhaps easier to see how the differencing operator works in the
special case of the difference $\MaxPoly((m,p),(0,0)) = p(m)$. I.e.,
if we have code that contains data of size $m$ and a program with
complexity $p$, then running the combination with no expectation of
remaining potential yields $p(m)$ available steps. The $\MaxPoly$ and
$\PlusPoly$ resource monoids will be used for the \ConsFree{} and
LFPL-style systems respectively, as we explain in
\autoref{sec:realising-iteration} and show how these resource monoids
yield the required polytime bounds on programs.

\paragraph{Agda Formalisation} The $\mathbb{N}$ resource monoid is
defined in \AgdaModule{Algebra.ResourceMonoid.Nat} and the polynomial
monoids are both defined in
\AgdaModule{Algebra.ResourceMonoid.Polynomial}. The definition is
parameterised by the ``size monoid'' operation (either $\sqcup$ or
$+$) used to compose sizes.

\subsubsection{Resource sub-monoids}

The separation between sizes of data and complexity of code in the
polynomial resource monoids motivates the use of resource sub-monoids
to ensure that programs themselves (as opposed to higher order code
which may contain closed over data) do not contain data that can be
iterated. We do this by requiring that programs' potential must come
from a specified resource sub-monoid:

\begin{definition}[Resource Sub-Monoids]
  A \emph{resource sub-monoid} $M_0 \subseteq M$ of a resource monoid
  $M$ consists of a subset $|M_0| \subseteq |M|$ that is closed under
  the monoid operations and $\mathit{acct}$.
\end{definition}

For both $\MaxPoly$ and $\PlusPoly$, the elements with zero size
component, i.e., of the form $(0,p)$, form a resource sub-monoid that
we will use for interpreting programs. We will call these sub-monoids
$\MaxPoly_0$ and $\PlusPoly_0$.

\subsection{Models of Quantitative Type Theory from Indexed Preorders}
\label{sec:qtt-models}

\citet{atkey18qtt} described a general class of QTT models termed
\emph{Quantitative Categories with Families}
(QCwFs). \citet{atkey18qtt} constructs QCwFs from certain Linear
Combinatory Algebras (LCAs), where terms in the $\sigma = 1$ fragment
are realised by elements of the LCA. However, there is a mistake in
that paper where the interpretation of contexts is stated to be the
category of \emph{assemblies} over the LCA, where it ought to be the
category of sets paired with realisability relations, with no
guarantee that all elements be realisable.

Here, we fix the mistake of \citet{atkey18qtt} and provide a more
general construction of QCwFs in terms of indexed linear preorders. We
construct indexed linear preorders specific to our polytime setting
below. They could also be constructed from LCAs.

\begin{definition}\label{def:indexed-linear-preorder}
  A \emph{$\mathbb{N}$-linear preorder}\footnote{We specialise to the semiring $\mathbb{N}$ here, but the same definition works for any suitable semiring $\mathcal{R}$.} is a preordered set
  $(L, \leq)$:
  \begin{enumerate}
  \item a commutative monoid $(I, -\otimes-)$ that is monotone
    w.r.t. the order;
  \item is closed: there is an operation
    $\multimap : L \times L \to L$ such that $x \otimes y \leq z$ iff
    $x \leq y \multimap z$; and
  \item has a function $\oc : \mathbb{N} \to L \to L$, to interpret
    resource requirement adjustments, satisfying:
    \begin{enumerate}
    \item $\oc_0 X \simeq I$, for discarding;
    \item $\oc_{m+n} X \leq (\oc_m X) \otimes (\oc_n X)$, for duplication;
    \item $\oc_m\oc_n X \leq \oc_{mn} X$ for nesting;
    \item $\oc_1 X \leq X$ for extraction / dereliction;
    \item $(\oc_n X) \otimes (\oc_n Y) \leq \oc_n(X \otimes Y)$, for distribution; and
    \item $n \leq m$ implies $\oc_n X \leq \oc_m X$, for usage weakening.
    \end{enumerate}
  \end{enumerate}
  The collection of all linear preorders and functions that preserve the
  order and the operations forms a category $\LinPreorder$.

  An \emph{indexed linear preorder} $L : \Set^\op \to \LinPreorder$ is
  a contravariant function, where we write $f^* : L(B) \to L(A)$ for
  the action of $L$ on functions $f : A \to B$, such that such that
  reindexing along projections has a right adjoint
  $L_{\Sigma_{a \in A}.B}(\pi_1^*X, Y) \cong L_{A}(X, \forall_B Y)$
  that commutes with reindexing.
\end{definition}

Given an indexed linear preorder $L : \Set^\op \to \LinPreorder$, we
construct a QCwF model of QTT with the basic type formers from
\autoref{sec:qtt-basic}:
\begin{enumerate}
\item Define a category $\cat{L}$ of interpretations of contexts with
  objects that are pairs $(A \in \Set, X \in L(A))$ and morphisms
  $f : (A,X) \to (B, Y)$ that are functions $f : A \to B$ such that
  $X \leq f^*Y$ (this is the Grothendieck category of $L$). There is a
  faithful functor $U : \cat{L} \to \Set$. The category $\cat{L}$ will
  be used for interpreting contexts in the $\sigma = 1$ fragment of
  QTT.
\item Define scaling of objects of $\cat{L}$ by
  $\pi(A,X) = (A, \oc_\pi X)$, and addition of $(A, X)$ and $(A, Y)$
  as $(A, X \otimes Y)$.
\item For each set $A$, define the collection of semantic types
  $\Ty(A)$ as the collection of $B : A \to \Set$ and
  $X \in L(\Sigma_{a \in A}.B(a))$. Thus a QTT type consists of an
  extensional meaning $B$ and its realisability specification $X$.
\item For each $A$ and $(B,X) \in \Ty(A)$, the $\sigma = 0$ fragment
  terms $\Tm(A,(B,X))$ are functions $\Pi_{a \in A}.\,B(a)$. For each
  context interpretation $(A, X)$ in $\cat{L}$ and type interpretation
  $(B,Y) \in \Ty(A)$, the $\sigma = 1$ fragment terms
  $\RTm((A,X),(B,Y))$ are functions $f : \Pi_{a \in A}.\,B(a)$ such
  that $X \leq \overline{f}^*Y$, where
  $\overline{f} : A \to \Sigma_{a \in A}.\,B(a)$ is the section
  associated with $f$.
\item The empty context is interpreted as $(\{*\}, I)$ and context
  extension $(A,X).n(B,Y)$ (i.e., comprehension) by
  $(\Sigma_{a \in A}.B(a), \pi_1^* X \otimes \oc_n Y)$.
\item Given $(A,X) \in \Ty(C)$ and
  $(B,Y) \in \Ty(\Sigma_{c \in C}.A(c))$, $\Sigma$-types are
  interpreted similarly to context extension and $\Pi$-types are
  interpreted as
  $(\lambda c.~(\Pi_{a \in A(c)}.\,B(c,a)), \forall_{A}(X \multimap
  (\mathit{ev}\,f)^*Y))$, where
  $\mathit{ev}\,f : (\Sigma_{c \in C}.A(c)) \to (\Sigma_{c \in
    C}.\Sigma_{a \in A(c)}.B(c,a))$ is defined using application of
  $f$.
\item Universe and Equality types are interpreted as normal in $\Set$
  with the realisability component set to $I$ in both cases. Note that
  the universe of small types includes resource-relevant realisability
  information for each type.
\item Realisability reflection for a type $(B,X) \in \Ty(A)$ is
  interpreted as the type
  $(\lambda a. \{ b \in B(a) \mid I \leq (\lambda a. (a,b))^*X \},
  I)$. Thus the set-component of the type is restricted to the
  elements that are realisable, while the actual realisability
  component is the ``empty'' $I$ realisability specification.
\end{enumerate}

\paragraph{Agda Formalisation} The indexed linear preorders are
defined in the Agda module \AgdaModule{IndexedLinear}. We have not yet
completed a formalisation of the construction of a full model of QTT
from an indexed linear preorder so this part is currently unmechanised.

\subsection{Amortised Complexity Realisability Model}
\label{sec:realisability-model}

Equipped with our underlying costed model of computation
(\autoref{sec:machine-model}) and a compositional notion of resource
potential (\autoref{sec:resource-monoids}), we can construct models of
QTT that witness the resource and type soundness of our complexity
constrained systems. We fix a resource monoid $M$ with sub-monoid
$M_0$ and proceed to build an indexed linear preorder of resource
accounted realisers.

\subsubsection{Indexed Linear Preorder}
\label{sec:realisability-indexed-linear-preorder}

We now define an indexed linear poset $L$ of realisers over $\Set$
that ties together our ``mathematical'' model of types in $\Set$ with
our machine model and resource monoid. This construction is a
reformulation of \citet{dallago11realisability}'s realisability models
to make it suitable for dependent types. For a set $A$, the carrier of
$L(A)$ is the set of ternary relations
$X \subseteq A \times M \times \ValSet$ and we define the ordering
$X \leq Y$ to hold iff there exists a realising expression
$E \in \ExpSet$ and potential $\gamma \in M_0$ such that for all
$a \in A$, $\alpha \in M$ and $v \in \ValSet$ with
$(a,\alpha,v) \in X$, we have that there exists a result
$v' \in \ValSet$, step count $k \in \mathbb{N}$ and result potential
$\beta \in M$ with:
\begin{enumerate}
\item $E, v \Downarrow_k v'$ (evaluation successfully completes in $k$ steps);
\item $(a, \beta, v') \in Y$ (the result is well-resourced and
  satisfies $Y$); and
\item $k \leq M(\alpha \rplus \gamma, \beta)$ (the step count is within the
  difference between the initial potential and the result potential).
\end{enumerate}
Note that the definition of realisablity is uniform in the element $a$
-- the realising expression $E$ and the potential $\gamma$ must work
for all $a$ -- thus the implementation and complexity measure of the
transition being modelled cannot depend on what the input is. Put in
implementation terms, the input $a$ is not present at
runtime. Moreover note that the potential $\gamma$ attached to the
expression $E$ must come from the sub-monoid $M_0$, indicating that is
intended to be data-free, while the potential $\alpha$ for the input
is from the full monoid $M$, so it can contain data and functions.

For $X, Y \in L(A)$, the required elements for symmetric monoidal
closed structure are defined as follows. For the tensor product
$X \otimes Y \in L(A)$, the realising value must be a pair $(v_1,v_2)$
and the potential of the pair must split into suitable potentials
$\alpha_1$, $\alpha_2$ for the components. For the residual
$X \multimap Y$, the realising value must be a closure with potential
to, when added to the potential of an input, compute the output with
enough remaining. Note that the potential attached to a closure
($\alpha$, here) need not be from the sub-monoid $M_0$. Unlike
top-level term interpretations, closures may contain data.
\begin{displaymath}
  \begin{array}{l@{\,}c@{\,}l}
    X \otimes Y &=& \{ (a, \alpha, (v_1, v_2)) \mid \exists \alpha_1, \alpha_2.~0 \leq M(\alpha, \alpha_1 \rplus \alpha_2) \land X(a,\alpha_1,v_1) \land Y(a,\alpha_2,v_2) \}\\
    X \multimap Y &=& \{ (a, \alpha, \clo{E}{\eta}) \mid
                      \begin{array}[t]{@{}l}
                        \forall \alpha' \in M, v,w \in \ValSet.\,X(a,\alpha',v) \Rightarrow\\
                        \quad \exists v', k, \beta.\,
                        E, (\eta :: w :: v) \Downarrow_k v' \land Y(a,\beta,v') \land k \leq M(\alpha \rplus \alpha', \beta) \} \end{array}
  \end{array}
\end{displaymath}
The seemingly useless $w \in \ValSet$ in the formula for
$X \multimap Y$ is a dummy argument standing for the self-referential
reference to the closure used for defining recursive programs.

Each $L(A)$ has a terminal (i.e. top) element, which is also the unit
for $\otimes$, defined as
$I_A = \{(a, \alpha, *) \mid a \in A, \alpha \in M\}$. The potential
$\alpha$ here is unrestricted, so $I_A$ can consume an arbitrary
resource.

$\mathbb{N}$-Graded exponentials in each $L(A)$ are defined using the
action of $(\mathbb{N}, \leq)$ on $M$ defined above. When $n > 0$, the
modality $\oc_n$ has no effect on realising values. It only serves to
alter the resource potentials. In the $n = 0$, case the realising
value must be $*$, in order to satisfy the $\oc_0 X \cong I$ condition
in \defref{def:indexed-linear-preorder} 3(a):
\begin{displaymath}
  \begin{array}{lcl}
    \oc_0\,X&=&\{(a, \alpha, *) \mid a \in A, \alpha \in M \} \\
    \oc_n\,X&=&\{(a,\alpha,v) \mid \exists \alpha'.\,M(n \cdot \alpha', \alpha) = 0 \land (a, \alpha', v) \in X \}
  \end{array}
\end{displaymath}

$L$ also has arbitrary $\Set$-indexed products, realised ``lazily'' as
functions that take dummy arguments. For $A \in \Set$ and
$B \in A \to \Set$ and $X \in L(\Sigma A.\,B)$, we define
$\forall_B X \in L(A)$ similarly to $\multimap$ above, but with
different resource and indexing requirements:
\begin{displaymath}
  \forall_B X = \{ (a,\alpha,\clo{E}{\eta}) \mid \forall b, v.~\exists v', \beta, k. E, (\eta :: v :: *) \Downarrow_k v' \land X((a,b),\beta,v') \land k \leq M(\alpha,\beta) \}
\end{displaymath}
Note, as with the definition of $X \leq Y$ above, the realiser closure
$\clo{E}{\eta}$ must be chosen uniformly for all $b$. This definition
also appears to allow arbitrary computation (paid for by $\alpha$) to
happen when the realising closure is applied, but the potential
$\alpha$ will only ever be greater than $\beta$ by enough to handle
the administrative costs of applying the function.

To complete the construction of $L$ as an indexed linear preorder, we
need to give realisers for each of the required inequalities in
\defref{def:indexed-linear-preorder}. In each case, this is a matter
of programming in the language of \autoref{sec:machine-model}. For
example, transitivity of the order is realised by sequencing of
expressions. The potentials are calculated by counting the steps in
the ensuing programs.

\begin{proposition}\label{prop:realisability-model}
  $L$, with $I$, $\otimes$, $\multimap$, $\oc_n$, and $\forall_{B}$
  defined above, is an indexed linear preorder.
\end{proposition}

\paragraph{Agda Formalisation} The construction of this indexed linear
preorder and the proof of \propref{prop:realisability-model} are
formalised in the Adga module
\AgdaModule{AmortisedRealisabilityModel}.

\subsubsection{Non-iterable Data Types}
\label{sec:noniter-model}

The model of QTT constructed in \propref{prop:realisability-model}
does not yet include any useful base types. Iterable types, which are
the ones that induce non-constant time complexities, require specific
properties of resource monoids that we introduce in
\autoref{sec:realising-iteration}.

Before that, we show how to define realisers for the representative
examples of non-iterable types from \autoref{sec:linear-case-analysis}
and \autoref{sec:noniter-qtt}. Booleans are the simplest case, with
only two cases and no chance of iteration. Lists are more complex: we
can have non-iterable lists containing iterable data.

\paragraph{Booleans} Fix $\mathbb{B} = \{ \mathit{tt}, \mathit{ff} \}$
as our set of boolean elements. We define an element of
$L(\mathbb{B})$ to represent boolean values:
\begin{displaymath}
  \mathrm{Bool} = \{ (\mathit{tt}, \alpha, \synTrue) \mid \alpha \in M \} \cup \{ (\mathit{ff}, \alpha, \synFalse) \mid \alpha \in M \}
\end{displaymath}
Thus, the boolean $\mathit{tt}$ is represented by the value $\synTrue$
and $\mathit{ff}$ is represented by $\synFalse$. In both cases, we
allow arbitrary potential $\alpha$ to be attached.

Realisability of the construction and elimination of booleans amounts
to the existence of the following inequalities. In any preorder
$L(A)$, we have $I_A \leq \mathit{tt}^* \mathrm{Bool}$ and
$I_A \leq \mathit{ff}^* \mathrm{Bool}$ (treating $\mathit{tt}$ and
$\mathit{ff}$ as constant functions $A \to \mathbb{B}$). These
inequalities are realised by the corresponding $\synTrue/\synFalse$
expression. For conditionals, the types involved are a little more
complex to ensure agreement between boolean manipulations at the
$\Set$-level and the realising computations. To get a realiser for a
conditional, we require a set $A$, an element $X \in L(A)$ (standing for
the context) and an element $Y \in L(A \times \mathbb{B})$ (standing
for the target type) and the existence in $L(A)$ of inequalities
$X \leq (\lambda a. (a, \mathit{tt}))^* Y$, for the true case, and
$X \leq (\lambda a. (a, \mathit{ff}))^* Y$, for the false case. When
we have all these, we get in $L(A \times \mathbb{B})$ an inequality
$\pi_1^* X \otimes \pi_2^* \mathrm{Bool} \leq Y$. This construction
suffices to realise the rules for QTT booleans in
\autoref{sec:noniter-qtt}.

\paragraph{Lists} Lists are a little more involved, due to the need to
explicitly manage a context that applies to all elements of the
list. Let $\mathrm{List}(B)$ be the set of lists with elements from a
set $B$. If we have $A : \Set$ and $B : A \to \Set$ and
$X \in L(\Sigma a:A.\, B a)$, then the resourced lists predicate
$\mathrm{RList}(X) \in L(\Sigma a : A.\, \mathrm{List}(B a))$ must
satisfy the equation:
\begin{displaymath}
  \mathrm{RList}(X) =
  \begin{array}[t]{l}
    \{ ((a, []), \alpha, (\synFalse, *)) \mid \alpha \in M \} \\
    \cup \\
    \{ \begin{array}[t]{@{}l}
         ((a, b :: bs), \alpha, (\synTrue, (v_1, v_2))) \mid \\
         \quad \exists \alpha_1, \alpha_2. 0 \leq M(\alpha, \alpha_1 \oplus \alpha_2) \land ((a, b), \alpha_1, v_1) \in X \land ((a, bs), \alpha_2, v_2) \in \mathrm{RList}(X) \}
       \end{array}
  \end{array}
\end{displaymath}
This equation has a least solution, by induction on the length of the
list being realised. This definition is somewhat involved, but in
essence states that a list is represented by tagged pairs, where
$\synFalse$ represents nil and $\synTrue$ represents cons, and that
the potential is distributed amongst the elements of the list as
needed.

\paragraph{Agda Formalisation} The construction of realisers for
booleans and lists are carried out in the Agda modules
\AgdaModule{AmortisedModel.Bool} and \AgdaModule{AmortisedModel.List}.

\section{Realising Iteration for Implicit Polynomial Time}
\label{sec:realising-iteration}

The models constructed in the previous section only allow for
constant-time programs to be realised. To interpret the iterators of
the \ConsFree{} and LFPL-style systems, we need to use the
$\MaxPoly$ and $\PlusPoly$ resource monoids. We do this in this
section, where first we establish some operations that will be useful
to see how they capture the nesting of iterations inherent to polytime
computation.

\subsection{Iteration Resource Monoids}

To interpret iteration over a resource monoid $(M, M_0)$, we require
additional structure, which we call an \emph{Iteration Resource
  Monoid} to account for measurement of the sizes of iterable data
structures and the effects of iteration on potentials.

\subsubsection{Definition}
We require:
\begin{enumerate}
\item a function $\mathit{size} : \mathbb{N} \to M$ that gives the
  potential of an iterable data structure of a given
  size;
\item a function $\mathit{raise} : M \to M$ that raises the
  (polynomial) degree of some potential; and
\item a function $\mathit{scale} : \mathbb{N} \times M \to M$ that
  scales a potential for a fixed number of iterations.
\end{enumerate}
These functions must satisfy the following properties:
\begin{enumerate}
\item $M_0$ is closed under the $\mathit{raise}$ operation;
\item for all $\alpha$ and $n$,
  $0 \leq M(\mathit{raise}(\alpha) \oplus \mathit{size}(n),
  \mathit{scale}(n,\alpha) \oplus \mathit{size}(n))$; and
\item for all $\alpha \in M_0$ and $n$,
  $0 \leq M(\mathit{scale}(1 + n, \alpha), \alpha \oplus
  \mathit{scale}(n, \alpha))$.
\end{enumerate}
The first property states that $\mathit{raise}$ is suitable as
potential for whole programs, meaning that it does not make any
requirements on the existence of iterable data. Note that we do
\emph{not} require $M_0$ to contain $\mathit{size}(n)$ -- programs
themselves may not contain iterable data, all potential for iteration
must be delivered externally. A useful intuition is that
$\mathit{scale}(n,\alpha)$ represents the potential required for at
most $n$ iterations that require potential $\alpha$, whereas
$\mathit{raise}(\alpha)$ represents the potential required for a
number of iterations that depends on the context. This is the
motivation behind the second required property, which states that
having $\mathit{raise}(\alpha)$ potential implies having
$\mathit{scale}(n,\alpha)$ potential when the current size is $n$. The
third property states that $\mathit{scale}$ decomposes as expected on
potentials that do not include any size potential.

Note that $\mathit{scale}(n,\alpha)$ is not the same as the action
$n \cdot \alpha$ defined in \autoref{sec:resource-monoids}. The latter
operation scales both size and function potential, but the former only
scales the function potential.

\subsubsection{Polynomial Iteration Resource Monoids}

Both of the polymonial resource monoids defined in
\defref{def:polynomial-resource-monoids} support the structure of an
Iteration Resource Monoid. We define:
\begin{enumerate}
\item $\mathit{size}(n) = (n, 0)$
\item $\mathit{raise}(n, p) = (n, xp)$
\item $\mathit{scale}(m, (n, p)) = (n, m \cdot p)$
\end{enumerate}
Note that $\mathit{raise}$ does indeed raise the degree of the
polynomial involved. Property 2 above is satisfied because for any
polynomial we have $(m \cdot p)(x) \leq (xp)(x)$ whenever $m \leq x$.

\subsubsection{Realising Iterable Natural Numbers}

For any natural number $n$, we define its representation as a value
$\mathrm{natValue}(n) \in \ValSet$ by recursion:
\begin{mathpar}
  \mathrm{natValue}(0) = (\synTrue , *)

  \mathrm{natValue}(1 + n) = (\synFalse, \mathrm{natValue}(n))
\end{mathpar}
This representation uses a tagged pair approach similar to our
representation of lists in \autoref{sec:noniter-model}. Using this, we
can define what it means for a natural number to be realisable via
$\mathrm{Nat} \in L(\mathbb{N})$:
\begin{displaymath}
  \mathrm{Nat} = \{(n, \alpha, \mathrm{natValue}(n)) \mid n \in \mathbb{N}, 0 \leq M(\alpha, \mathit{size}(n+1)) \}
\end{displaymath}
So a natural number $n$ is realised by the value
$\mathrm{natValue}(n)$ as long as we have at least
$\mathit{size}(n+1)$ potential (we add one to make the LFPL soundness
proof easier).  This gives us the ability to represent natural numbers
as a type in QTT, but in order to iterate (and construct in the case
of LFPL), we need to construct specific realisers for the
\ConsFree{} and LFPL systems.

\subsection{The Cons-Free System}
\label{sec:consfree-sound}

The \ConsFree{} system uses the $\MaxPoly$ resource monoid, with
the distinguished sub-monoid being those elements that are $0$ in the
size component. We enumerate the features of the \ConsFree{}
system and justify their realisability with the $\MaxPoly$ resource
monoid:
\begin{enumerate}
\item Duplication of natural numbers by $\dupNat(M)$ is realisable by
  the expression $(0,0)$, which creates a pair by copying the input
  variable twice. By the cost semantics in
  \autoref{sec:machine-model}, this takes $1$ step of computation (we
  assume that it is actually implemented via some pointer copy).  The
  resource accounting for this realiser works because the size
  component required for the output is the maximum of the size
  components of the two elements, and since $n \sqcup n = n$, we have
  enough resources to fulfil this.
\item Construction of natural numbers is not realisable. In a putative
  $\conSucc$ rule, we would need to get an additional unit of size
  resource from nowhere.
\item Iteration is realised by constructing a realising expression in
  the expression language from the given expressions for the zero and
  successor cases that uses the in-built recursion of the
  language. The proof that resources are correctly accounted for is
  carried out by induction on the natural being iterated over. For
  $n$, we require potential
  $\mathit{scale}(n,\mathit{acct}(4) \rplus \gamma_{\mathit{succ}})
  \rplus (\mathit{acct}(2) \rplus \gamma_{\mathit{zero}})$, where
  $\gamma_{\mathit{succ}}$ and $\gamma_{\mathit{zero}}$ are the
  potentials required by the successor and zero cases respectively. By
  Property (2) of iteration resource monoids, above, we know that
  $\mathit{raise}(\mathit{acct}(4) \rplus \gamma_{\mathit{succ}})
  \rplus (\mathit{acct}(2) \rplus \gamma_{\mathit{zero}})$ always
  dominates this requirement when paired with the potential
  $\mathit{size}(n)$ from the input. Therefore, this latter
  expression, plus some administrative set up costs, is the required
  potential for the whole iterator.
\end{enumerate}
Together, we have a soundness result for the \ConsFree{} system,
that ensures that every term in the $\sigma = 1$ fragment is
realisable by a \emph{correct} program that terminates in polynomial
time for all inputs:
\begin{theorem}[Soundness for the \ConsFree{} System]
  \label{thm:cons-free-soundness}
  If we have a term $n \stackrel1: \tyNat \vdash M \stackrel1: T(n)$ then there
  exists a realising expression $E$ and polynomial $p$ such that for
  all $n \in \mathbb{N}$, there exists $v \in \ValSet$ and
  $k \in \mathbb{N}$ such that
  $E, [\mathrm{natValue}(n)] \Downarrow_k v$, $k \leq p(n)$ and $v$ is
  a realising value for
  $\llbracket M \rrbracket(n) \in \llbracket T \rrbracket(n)$.
\end{theorem}

\paragraph{Agda Formalisation} The realisability of the
\ConsFree{} system iterator and the soundness property of the
whole system are formalised in the Agda modules \AgdaModule{ConsFree}
and \AgdaModule{ConsFree.Iterator}. The soundness theorem is a
combination of this and the QTT model sketched in
\autoref{sec:qtt-models}.

\subsection{The LFPL System}
\label{sec:lfpl-sound}

The LFPL system uses the $\PlusPoly$ resource monoid, with the
distinguished sub-monoid again being those elements that are $0$ in
the size component. With this resource monoid, the capabilities
offered at the QTT level are altered:
\begin{enumerate}
\item We can no longer duplicate natural numbers, because
  $\tyNat \otimes \tyNat$ requires twice as much size resource as
  $\tyNat$, due to the combining operation on size potentials being
  addition.
\item We define the realisability specification for diamonds
  $\Diamond \in L(1)$ to be
  $\Diamond = \{ (*, \alpha, *) \mid 0 \leq M(\alpha,
  \mathit{size}(1)) \}$.  Thus, a diamond represents at least one unit
  of size resource, matching the intuitive explanation given in
  \autoref{sec:lfpl-intro}.
\item With this definition of realisability for $\Diamond$s, it is
  possible to realise the $\conZero$ and $\conSucc$ constructors for
  natural numbers. By the additive combination of size resources we
  get $1$ from the diamond and $n+1$ from the predecessor to total
  $n+2$ for a new number. Note that, even if we add a $\Diamond$ type
  to the \ConsFree{} system, it would still not be possible to
  realise the constructors, because we would only have
  $1 \sqcup (n+1) = n+1$ size resource for the output.
\item The construction of the recursor follows a very similar proof to
  the realisability of \ConsFree{} iterator, up to some
  additional work to make sure that the dummy $*$ values representing
  the diamond components end up in the right places. This additional
  work is revealed in the required potential for the LFPL iterator
  being
  $\mathit{raise}(\mathit{acct}(8) \rplus \gamma_{\mathit{succ}})
  \rplus (\mathit{acct}(2) \rplus \gamma_{\mathit{zero}})$, so
  slightly higher in the successor case.
\end{enumerate}

Soundness for the LFPL system is similar to the \ConsFree{}
system, except for a $+1$ to the input to the polynomial, to account
for the fact that we cost one size unit for the $\conZero$
constructor.
\begin{theorem}[Soundness for the LFPL-style System]
  \label{thm:lfpl-soundness}
  If we have a term $n \stackrel1: \tyNat \vdash M \stackrel1: T(n)$ then there
  exists a realising expression $E$ and polynomial $p$ such that for
  all $n \in \mathbb{N}$, there exists $v \in \ValSet$ and
  $k \in \mathbb{N}$ such that
  $E, [\mathrm{natValue}(n)] \Downarrow_k v$, $k \leq p(n+1)$ and $v$
  is a realising value for
  $\llbracket M \rrbracket(n) \in \llbracket T \rrbracket(n)$.
\end{theorem}

The proofs of well-accounted realisability for the LFPL iterator, and
the \ConsFree{} iterator, could be adapted to any other
inductively defined type that is finitely branching. This is not
immediately necessary, as evidenced by the construction of other
datatypes in \autoref{sec:prog-datatypes}. Nevertheless, native tree
type where the iterability is proportional to the total number of
nodes would be useful.

\paragraph{Agda Formalisation} The realisability of the LFPL system
iterator and the soundness property of the whole system are formalised
in the Agda modules \AgdaModule{LFPL} and \AgdaModule{LFPL.Iterator}.

\section{Related and Future Work}
\label{sec:conclusion}

We have presented two extensions of Quantitative Type Theory that
soundly and completely capture polynomial time. This allows for an
expressive combination of verification and complexity constrained
computation, including characterisations of the classes P, NP, and
BPP. We now discuss related work, and take a look at where the
combination of polytime and dependency could take us.

\subsection{Related Work}

\paragraph{Implicit Computational Complexity with Linear Types}
Implicit Computational Complexity \cite{Lago11} is a vast field, so we
only survey closely related works. We have already mentioned the
Bounded Linear Logic \cite{bll92}, Soft Affine Logic \cite{Lafont04},
Light Linear Logic \cite{lll98} and LFPL \cite{hofmann99lfpl} systems,
which all use linear typing to implicitly capture polynomial
time. \citet{Jones01} characterises polynomial time using first-order
functional programs without constructors. Thus it shares a method with
our \ConsFree{} system, but we use linear typing to permit
controlled use of higher-order functions. Other approaches to
polynomial time use stratification or information flow tracking to
ensure that the outputs of iteration may not be used unrestrictedly to
drive further iteration. For example, \cite{BellantoniC92} and
\cite{HainryP23}. Below polynomial time, systems have be devised to
capture LOGSPACE \cite{LagoS16}. Above polynomial time, systems such
as Elementary Affine Logic (EAL) capture all Elementary-time functions
\cite{CoppolaM01}.

We have used \citet{dallago11realisability}'s technique to prove
soundness of our extension of QTT. This technique has been
successfully applied to many other linear typing based systems, such
as BLL \cite{hofmann04bll-realisability,dallago10bll} and LLL
\cite{LagoH10} and EAL. In contrast to most of those systems, we do
not use restricted $\oc$-modalities and second order encodings to
express datatypes. Our explicit datatype approaches enabled our
combination of dependent types and polynomial time.

\paragraph{Explicit Resource Accounting with Dependent Types} In
contrast to the implicit systems, previous works have constructed
systems that give explicit resource bounds via typing. Examples
include \citet{HoffmannDW17}'s Resource Allocated ML (RAML) and
\citet{RajaniG0021}, both of which are based on ideas of type-based
amortised complexity analysis arising from \citet{hofmann99lfpl}'s
ideas, via the work of \citet{HofmannJ03}. More details are to be
found in the survey paper of \citet{HoffmannJ22}. Another approach is
to track costs at the value level instead of the
types. \citet{Danielsson08} describes a system that uses a ``tick''
effect to count steps of computation, which can be reasoned about via
dependent types. \citet{NiuSGH22} take this idea further by employing
a modality-based phase separation to ensure that tick counting never
interferes with the functional business of
programs. \citet{McCarthyFNFF16} is another tick effect based system
in Coq. All of these tick-counting techniques rely on the programmer
correctly annotating the program with tick effects to count the
resource usage they are interested in, in contrast our intrinsic
approach.

\paragraph{Linear and Substructural Dependent Types} We chose QTT as
the particular combination of linear and dependent types for our
systems. Other systems include systems such as those by
\citet{CervesatoP02}, \citet{KrishnaswamiPB15}, and \citet{Vakar14}
which all use a strict separation between linear and non-linear
variables. This strict separation would mean that we could not as
easily move programs from the linear fragment into the types, as in
\autoref{sec:programming-polytime}. Systems that are more like QTT in
that they do not have a strict separation of variables include those
of \citet{MoonEO21}, \citet{ChoudhuryEEW21}, and \citet{Abel23}. These
systems differ from QTT in that they do not include a complete copy of
unrestricted type theory as QTT does in its $\sigma = 0$ fragment,
because they all track usage in types as well as terms, so it is not
clear how to use them for unrestricted reasoning as we do with
QTT. \citet{FuKS22} present a system that is closer to QTT but does
not include a universe type, which we used in \autoref{sec:classes} to
be able to characterise complexity classes as predicates decidable in
restricted complexity.

\subsection{Future Work}
\label{sec:towards-synthetic-complexity-theory}

\paragraph{Implementation} We currently lack an implementation of our
extension of QTT, which hampers further investigation of programming
and proving with polytime along the lines of
\autoref{sec:programming-polytime}. Idris 2 \cite{Brady21} is an
implementation of QTT, but cannot be used directly because its facility
for defining datatypes is too liberal, not making a distinction
between iterable and non-iterable datatypes. A further
implementation-focused question is whether or not the term-level
polytime guarantees can be turned to type-level guarantees to
guarantee polytime typechecking.

\paragraph{Other Complexity Classes}
We have been able to characterise the classes NP and BPP in terms of
our underlying characterisation of P (\autoref{sec:classes}). It seems
straighforward to extend this to related classes like coNP, RP,
etc. It also seems feasible to adapt the techniques presented here to
other complexity classes such as LOGSPACE and ELEMENTARY, given the
simply typed linear systems mentioned above. Complexity classes based
on circuits may be more challenging, but we do now have a way to
characterise circuits that are generatable in polynomial
time.

\paragraph{Explicit Resource Tracking}
Our construction already includes soundness of a system with intrinsic
but explicit resource tracking where $\Diamond$s are used to pay for
every step of computation but never returned, via the natural number
resource monoid defined in
\autoref{sec:specific-monoids}. Investigation of such a system may
yield a system that tracks the intrinsic cost of programs precisely
and explicitly.

\paragraph{Towards a Synthetic Computational Complexity Theory?}

The realisability type $\mathbf{R}(A)$ described in
\autoref{sec:qtt-reflection} allows us to internalise the
realisability of certain functions into the logical ($\sigma = 0$)
fragment of the calculus. However, it is not possible to derive any
logical consequences from this other than turning it back into a
function. This limitation becomes acute when trying to prove results
from standard Computational Complexity theory. Even though we can
characterise the class NP, as we did in \autoref{sec:np-class}, and it
is a ``matter of programming'' to show that 3-SAT is in NP, we cannot
prove the Cook-Levin theorem that 3-SAT is NP-complete. This is
because the proof relies on obtaining the \emph{source code} of the
program solving an NP problem and then encoding that program in
3-SAT. To do this in our setting, we would need to internalise the
soundness property (\thmref{thm:lfpl-soundness}) as an axiom, stating
that for a realisable polytime function there (merely) exists a
realising expression $E$ that completes in polynomial time, and then
writing polytime encodings into 3-SAT. We hope that the addition of
such an axiom to our system would lead to an expressive machine-free
\emph{Synthetic Computational Complexity Theory}, analogous to the
Church-Turing axiom for Synthetic Computability Theory as described by
\citet{Bauer06}.

\begin{acks}
  Thanks to Anton Lorenzen and Fredrik Nordvall Forsberg for their
  comments on an earlier revision, and to the anonymous POPL reviewers
  for their detailed, perceptive, and interesting reviews. This work
  was funded by the \grantsponsor{EPSRC}{Engineering and Physical
    Sciences Research
    Council}{https://www.ukri.org/about-us/epsrc/}: Grant number
  \grantnum{EPSRC}{EP/T026960/1}, \emph{AISEC: AI Secure and
    Explainable by Construction}.
\end{acks}

\section*{Data Availability Statement}

The Agda source files and rendered HTML for this paper is available
from Zenodo \cite{atkey2023polydep-artefact}. The source files are
also available online at GitHub:
\url{https://github.com/bobatkey/qtt-models}.

\bibliographystyle{ACM-Reference-Format}
\bibliography{bibliography}

\input{appendix-rules.tex}

\end{document}

%% file: appendix-rules.tex
\appendix

\section{Typing Rules for QTT/Cons-free and QTT/LFPL}
\label{sec:rules}

The judgements of Quantitative Type Theory are as follows:
\begin{mathpar}
  \begin{array}{ll}
    \Gamma~\isctxt&\textrm{contexts} \\
    \Gamma \vdash S~\istype&\textrm{types} \\
    \Gamma \vdash S \equiv T~\istype&\textrm{equal types} \\
  \end{array}

  \begin{array}{ll}
    \Gamma \vdash M \stackrel\sigma: S&\textrm{terms} \\
    \Gamma \vdash M \equiv N \stackrel\sigma: S&\textrm{equal terms}
  \end{array}
\end{mathpar}
In the term and term equality judgements, the usage $\sigma$ is either
$0$ or $1$. We use $\rho$ and $\pi$ to range over arbitrary usages
from the semiring $R$.

\subsection{Context formation}
\begin{mathpar}
  \inferrule* [right=Emp]
  { }
  {\diamond~\isctxt}

  \inferrule* [right=Ext]
  {\Gamma~\isctxt \\ 0\Gamma \vdash S}
  {\Gamma, x \stackrel\rho: S~\isctxt}
\end{mathpar}

\subsection{Type Equality}
\begin{mathpar}
  \inferrule* [right=Ty-Eq-Refl]
  {0\Gamma \vdash S}
  {0\Gamma \vdash S \equiv S}

  \inferrule* [right=Ty-Eq-Symm]
  {0\Gamma \vdash S \equiv T}
  {0\Gamma \vdash T \equiv S}

  \inferrule* [right=Ty-Eq-Tran]
  {0\Gamma \vdash S \equiv T \\ 0\Gamma \vdash T \equiv U}
  {0\Gamma \vdash S \equiv U}
\end{mathpar}
as well as congruence rules for each type formation rule (elided), and
the universe eliminator.

\subsection{Term Equality}
\begin{mathpar}
  \inferrule* [right=Tm-Eq-Refl]
  {\Gamma \vdash M \stackrel\sigma: S}
  {\Gamma \vdash M \equiv M \stackrel\sigma: S}

  \inferrule* [right=Tm-Eq-Symm]
  {\Gamma \vdash M \equiv N \stackrel\sigma: S}
  {\Gamma \vdash N \equiv M \stackrel\sigma: S}

  \inferrule* [right=Tm-Eq-Tran]
  {\Gamma \vdash M \equiv N \stackrel\sigma: S \\
    \Gamma \vdash N \equiv O \stackrel\sigma: S}
  {\Gamma \vdash M \equiv O \stackrel\sigma: S}
\end{mathpar}
as well as congruence rules for each term formation rule (elided), and
the specific $\beta\eta$-equalities for each type listed below.

\subsection{Variables, conversion, sub-usaging}
\begin{mathpar}
  \inferrule* [right=Var]
  {0\Gamma, x \stackrel\sigma: S, 0\Gamma'~\isctxt}
  {0\Gamma, x \stackrel\sigma: S, 0\Gamma' \vdash x \stackrel\sigma: S}

  \inferrule* [right=Conv]
  {\Gamma \vdash M \stackrel\sigma: S \\ 0\Gamma \vdash S \equiv T~\istype}
  {\Gamma \vdash M \stackrel\sigma: T}

  \inferrule* [right=Sub]
  {\Gamma \vdash M \stackrel\sigma: S \\ \Gamma' \sqsubseteq \Gamma}
  {\Gamma' \vdash M \stackrel\sigma: S}
\end{mathpar}

\subsection{$\Pi$-types} Type formation:
\begin{displaymath}
  \inferrule* [right=Ty-Pi]
  {0\Gamma \vdash S \\ 0\Gamma, x \stackrel0: S \vdash T}
  {0\Gamma \vdash (x \stackrel\pi: S) \to T}
\end{displaymath}
Introduction and elimination:
\begin{mathpar}
  \inferrule* [right=Tm-Lam]
  {\Gamma, x \stackrel{\sigma\pi}: S \vdash M \stackrel\sigma: T}
  {\Gamma \vdash \lambda x.M \stackrel\sigma: (x \stackrel\pi: S) \to T}

  \inferrule* [right=Tm-App]
  {\Gamma_1 \vdash M \stackrel\sigma: (x \stackrel\pi: S) \to T \\
    \Gamma_2 \vdash N \stackrel{\sigma'}: S \\
    0\Gamma_1 = 0\Gamma_2 \\
    \sigma' = 0 \Leftrightarrow (\pi = 0 \lor \sigma = 0)
  }
  {\Gamma_1 + \pi\Gamma_2 \vdash M\,N \stackrel\sigma: T[N/x]}
\end{mathpar}
$\beta\eta$-equalities (as well as congruences):
\begin{mathpar}
  \inferrule* [right=Tm-Eq-Pi$\beta$]
  {\Gamma_1, x \stackrel{\sigma\pi}: S \vdash M \stackrel\sigma: T \\
    \Gamma_2 \vdash N \stackrel{\sigma'}: S \\
    (\sigma' = 0 \Leftrightarrow \pi = 0 \lor \sigma = 0)}
  {\Gamma_1 + \pi\Gamma_2 \vdash (\lambda x.M)\, N \equiv M[N/x] \stackrel\sigma: T[N/x]}

  \inferrule* [right=Tm-Eq-Pi$\eta$]
  {\Gamma \vdash M \stackrel\sigma: (x \stackrel\pi: S) \to T}
  {\Gamma \vdash \lambda x .M\,x \equiv M \stackrel\sigma: (x \stackrel\pi: S) \to T}
\end{mathpar}

\subsection{$\Sigma$-types}
Type formation:
\begin{mathpar}
  \inferrule* [right=Ty-Tensor]
  {0\Gamma \vdash S~\istype \\ 0\Gamma, x \stackrel0: S \vdash T~\istype}
  {0\Gamma \vdash (x \stackrel\pi: S) \otimes T~\istype}

  \inferrule* [right=Ty-Unit]
  {0\Gamma~\isctxt}
  {0\Gamma \vdash I~\istype}
\end{mathpar}
Introduction:
\begin{mathpar}
  \inferrule* [right=Tm-Pair]
  {\Gamma_1 \vdash M \stackrel{\sigma'}: S \\
    \Gamma_2 \vdash N \stackrel\sigma: T[M/x] \\
    0\Gamma_1 = 0\Gamma_2 \\
    \sigma' = 0 \Leftrightarrow (\pi = 0 \lor \sigma = 0)}
  {\pi\Gamma_1 + \Gamma_2 \vdash (M, N) \stackrel\sigma: (x \stackrel\pi: S) \otimes T}

  \inferrule* [right=Tm-Unit]
  {0\Gamma~\isctxt}
  {0\Gamma \vdash * \stackrel\sigma: I}
\end{mathpar}
$\sigma = 0$ fragment eliminators:
\begin{mathpar}
  \inferrule* [right=Tm-Fst]
  {\Gamma \vdash M \stackrel0: (x \stackrel\pi: S) \otimes T}
  {\Gamma \vdash \mathrm{fst}(M) \stackrel0: S}

  \inferrule* [right=Tm-Snd]
  {\Gamma \vdash M \stackrel0: (x \stackrel\pi: S) \otimes T}
  {\Gamma \vdash \mathrm{snd}(M) \stackrel0: T[\mathrm{fst}(M)/x]}
\end{mathpar}
Linear eliminators:
\begin{mathpar}
  \inferrule* [right=Tm-Let-Pair]
  {0\Gamma, z \stackrel0: (x \stackrel\pi: A) \otimes B \vdash C~\istype \\
    \Gamma_1 \vdash M \stackrel\sigma: (x \stackrel\pi: A) \otimes B \\
    \Gamma_2, x \stackrel{\sigma\pi}: A, y \stackrel{\sigma}: B \vdash N \stackrel\sigma: C[(x,y)/z]\\
    0\Gamma_1 = 0\Gamma_2}
  {\Gamma_1 + \Gamma_2 \vdash \mathrm{let}~(x,y) = M~\mathrm{in}~N \stackrel\sigma: C[M/z]}

  \inferrule* [right=Tm-Let-Unit]
  {0\Gamma_1, x \stackrel0: I \vdash C~\istype \\
    \Gamma_1 \vdash M \stackrel\sigma: I \\
    \Gamma_2 \vdash N \stackrel\sigma: C[*/x] \\
    0\Gamma_1 = 0\Gamma_2}
  {\Gamma_1+\Gamma_2 \vdash \mathrm{let}~* = M~\mathrm{in}~N \stackrel\sigma: C[M/x]}
\end{mathpar}
$\beta$-equality for both fragments:
\begin{mathpar}
  \inferrule*
  {0\Gamma, z \stackrel0: (x \stackrel\pi: A) \otimes B \vdash C~\istype \\
    \Gamma_1 \vdash M_1 \stackrel{\sigma'}: A \\
    \Gamma_2 \vdash M_2 \stackrel\sigma: B[M/x] \\
    \sigma' = 0 \Leftrightarrow (\pi = 0 \lor \sigma = 0) \\
    \Gamma_3, x \stackrel{\sigma\pi}: A, y \stackrel{\sigma}: B \vdash N \stackrel\sigma: C[(x,y)/z]\\
    0\Gamma_1 = 0\Gamma_2 = 0\Gamma_3}
  {\pi\Gamma_1 + \Gamma_2 + \Gamma_3 \vdash \mathrm{let}~(x,y) = (M_1,M_2)~\mathrm{in}~N \equiv N[M_1/x,M_2/y]\stackrel\sigma: C[(M_1,M_2)/z]}

  \inferrule*
  {0\Gamma, x \stackrel0: I \vdash C~\istype \\
    \Gamma \vdash N \stackrel\sigma: C[*/x]}
  {\Gamma \vdash \mathrm{let}~* = *~\mathrm{in}~N \equiv N[*/x] \stackrel\sigma: C[*/x]}
\end{mathpar}
$\beta\eta$-equalities for the $\sigma = 0$ fragment:
\begin{mathpar}
  \inferrule* [right=Eq-Pair-Fst]
  {0\Gamma \vdash M \stackrel0: S \\
    0\Gamma \vdash N \stackrel0: T[M/x]}
  {0\Gamma \vdash \mathrm{fst}(M, N) \equiv M \stackrel0: S}

  \inferrule* [right=Eq-Pair-Snd]
  {0\Gamma \vdash M \stackrel0: S \\
    0\Gamma \vdash N \stackrel0: T[M/x]}
  {0\Gamma \vdash \mathrm{snd}(M, N) \equiv N \stackrel0: T[M/x]}

  \inferrule* [right=Eq-Unit-$\eta$]
  {0\Gamma \vdash M \stackrel0: I}
  {0\Gamma \vdash M \equiv * \stackrel0: I}

  \inferrule* [right=Eq-Pair-$\eta$]
  {0\Gamma \vdash M \stackrel0: (x \stackrel\pi: S) \otimes T}
  {0\Gamma \vdash (\mathrm{fst}(M), \mathrm{snd}(M)) \equiv M \stackrel0: (x \stackrel\pi: S) \otimes T}

  \inferrule*
  {0\Gamma, z \stackrel0: (x \stackrel\pi: A) \otimes B \vdash C \\
    0\Gamma \vdash M \stackrel0: (x \stackrel\pi: A) \otimes B \\
    0\Gamma, x \stackrel0: A, y \stackrel0: B \vdash N \stackrel0: C[(x,y)/z]}
  {0\Gamma \vdash \mathrm{let}~(x,y) = M~\mathrm{in}~N \equiv N[\mathrm{fst}(M)/x,\mathrm{snd}(M)/y] \stackrel0: C[M/z]}
\end{mathpar}

\subsection{Identity Type} Type formation, introduction, reflection
and $\eta$-law:
\begin{mathpar}
  \inferrule* [right=Ty-Id]
  {0\Gamma \vdash S~\istype \\
    0\Gamma \vdash M \stackrel0: S \\
    0\Gamma \vdash N \stackrel0: S}
  {0\Gamma \vdash M =_S N~\istype}

  \inferrule* [right=Id-Refl]
  {\Gamma \vdash M \stackrel\sigma: S}
  {\Gamma \vdash \conRefl(M) \stackrel\sigma: M =_S M}

  \inferrule* [right=Id-Reflect]
  {\Gamma \vdash N \stackrel0: M_1 =_S M_2}
  {\Gamma \vdash M_1 \equiv M_2 \stackrel0: S}

  \inferrule* [right=Id-Uniq]
  {0\Gamma \vdash P \stackrel0: M =_A M}
  {0\Gamma \vdash P \equiv \mathrm{refl}(M) : M=_A M}
\end{mathpar}

\subsection{Universe}

\newcommand{\TySet}{\mathsf{U}}
\newcommand{\TyEl}{\mathsf{El}}
Type formation:
\begin{displaymath}
  \inferrule* [right=Ty-U]
  {0\Gamma~\isctxt}
  {0\Gamma \vdash \TySet~\istype}
\end{displaymath}
Introduction (also with introduction rules for all other type formers except $\TySet$):
\begin{mathpar}
  \inferrule* [right=Tm-U-Bool]
  {\Gamma~\isctxt}
  {\Gamma \vdash \BoolTy \stackrel\sigma: \TySet}

  \inferrule* [right=Tm-U-Tensor]
  {\Gamma \vdash M \stackrel\sigma: \TySet \\
    \Gamma, x \stackrel\rho: \TyEl(M) \vdash N \stackrel\sigma: \TySet}
  {\Gamma \vdash (x \stackrel\pi: M) \otimes N \stackrel\sigma: \TySet}

  \ldots
\end{mathpar}
Elimination:
\begin{displaymath}
  \inferrule* [right=Ty-El]
  {0\Gamma \vdash M \stackrel0: \TySet}
  {0\Gamma \vdash \TyEl(M)~\istype}
\end{displaymath}
Equality:
\begin{mathpar}



  \inferrule* [right=Ty-Eq-El-Cong]
  {0\Gamma \vdash M \equiv N \stackrel0: \TySet}
  {0\Gamma \vdash \TyEl(M) \equiv \TyEl(N)~\istype}
\end{mathpar}

\subsection{Booleans}
Formation, introduction, and elimination:
\begin{mathpar}
  \inferrule*
  {\Gamma~\isctxt}
  {\Gamma \vdash \BoolTy~\istype}

  \inferrule*
  {\Gamma~\isctxt}
  {0\Gamma \vdash \cTrue, \cFalse \stackrel\sigma: \BoolTy}

  \inferrule*
  {0\Gamma_1, x \stackrel0: \BoolTy \vdash P~\istype \\
    \Gamma_1 \vdash M \stackrel\sigma: \BoolTy \\
    \Gamma_2 \vdash N_t \stackrel\sigma: P[\cTrue/x] \\
    \Gamma_2 \vdash N_f \stackrel\sigma: P[\cFalse/x] \\
    0\Gamma_1 = 0\Gamma_2}
  {\Gamma_1 + \Gamma_2 \vdash \If_{x.P}\: M \: \Then \: N_t \: \Else \: N_f \stackrel\sigma: P[M/x]}
\end{mathpar}
$\beta$-equalities:
\begin{mathpar}
  \inferrule* [right=Tm-Eq-True$\beta$]
  {0\Gamma, z \stackrel0: \BoolTy \vdash P \\
    \Gamma \vdash M_t \stackrel\sigma: P[\mathrm{true}/z] \\
    \Gamma \vdash M_f \stackrel\sigma: P[\mathrm{false}/z]}
  {\Gamma \vdash \If_{x.P}\: \cTrue \: \Then \: N_t \: \Else \: N_f \equiv N_t \stackrel\sigma: P[\mathrm{true}/z]}

  \inferrule* [right=Tm-Eq-False$\beta$]
  {0\Gamma, z \stackrel0: \BoolTy \vdash P \\
    \Gamma \vdash M_t \stackrel\sigma: P[\mathrm{true}/z] \\
    \Gamma \vdash M_f \stackrel\sigma: P[\mathrm{false}/z]}
  {\Gamma \vdash \If_{x.P}\: \cFalse \: \Then \: N_t \: \Else \: N_f \equiv N_f \stackrel\sigma: P[\mathrm{false}/z]}
\end{mathpar}

\subsection{Lists} Formation and introduction:
\begin{mathpar}
  \inferrule*
  {0\Gamma \vdash T~\istype}
  {0\Gamma \vdash \ListTy(T)~\istype}

  \inferrule*
  {\Gamma \vdash T~\istype}
  {0\Gamma \vdash \cNil \stackrel\sigma: \ListTy(T)}

  \inferrule*
  {\Gamma_1 \vdash M \stackrel\sigma: T \\
    \Gamma_2 \vdash N \stackrel\sigma: \ListTy(T) \\
    0\Gamma_1 = 0\Gamma_2}
  {\Gamma_1 + \Gamma_2 \vdash \cCons(M,N) \stackrel\sigma: \ListTy(T)}
\end{mathpar}
Case analysis:
\begin{mathpar}
  \inferrule*
  {0\Gamma_1, x \stackrel0: \ListTy(T) \vdash P~\istype \\
    \Gamma_1 \vdash M \stackrel\sigma: \ListTy(T) \\
    \Gamma_2 \vdash N_1 \stackrel\sigma: P[\cNil/x] \\
    \Gamma_2, h \stackrel\sigma: T, t \stackrel\sigma: \ListTy(T) \vdash N_2 \stackrel\sigma: P[\cCons(h,t)/x] \\
    0\Gamma_1 = 0\Gamma_2}
  {\Gamma_1 + \Gamma_2 \vdash \Match_{x.P}\,M\,\{\,\cNil \mapsto N_1; \cCons(h,t) \mapsto N_2\,\} \stackrel\sigma: P[M/x]}
\end{mathpar}
$\beta$-equalities for case analysis:
\begin{mathpar}
  \inferrule* [right=Eq-List-Match-Nil]
  {0\Gamma, x : \stackrel0: \ListTy(T) \vdash P~\istype \\
    \Gamma \vdash N_1 \stackrel\sigma: P[\cNil/x] \\
    \Gamma, h \stackrel\sigma: T, t \stackrel\sigma: \ListTy(T) \vdash N_2 \stackrel\sigma: P[\cCons(h,t)/x]}
  {\Gamma \vdash \Match_{x.P}\,\cNil\,\{\,\cNil \mapsto N_1; \cCons(h,t) \mapsto N_2\,\} \equiv N_1 \stackrel\sigma: P[\cNil/x]}

  \inferrule* [right=Eq-List-Match-Cons]
  {0\Gamma_1, x : \stackrel0: \ListTy(T) \vdash P~\istype \\
    \Gamma_1 \vdash M_1 \stackrel\sigma: T \\
    \Gamma_2 \vdash M_2 \stackrel\sigma: \ListTy(T) \\
    \Gamma_3 \vdash N_1 \stackrel\sigma: P[\cNil/x] \\
    \Gamma_3, h \stackrel\sigma: T, t \stackrel\sigma: \ListTy(T) \vdash N_2 \stackrel\sigma: P[\cCons(h,t)/x] \\
    0\Gamma_1 = 0\Gamma_2 = 0\Gamma_3}
  {\Gamma_1 + \Gamma_2 + \Gamma_3 \vdash {\begin{array}[m]{@{}l}
                                            \Match_{x.P}\,(\cCons(M_1,M_2))\\
                                            \qquad\{\,\cNil \mapsto N_1; \cCons(h,t) \mapsto N_2\,\}\\
                                            \equiv N_2[M_1/h,M_2/t] \end{array}} \stackrel\sigma: P[\cCons(M_1,M_2)/x]}
\end{mathpar}
$\sigma = 0$ recursive eliminator:
\begin{mathpar}
  \inferrule*
  {0\Gamma, x \stackrel0: \ListTy(T) \vdash P~\istype \\
    0\Gamma \vdash M \stackrel0: \ListTy(T) \\
    0\Gamma \vdash N_1 \stackrel0: P[\cNil/x] \\
    0\Gamma, h \stackrel0: T, t \stackrel0: \ListTy(T), p \stackrel0: P[t/x] \vdash N_2 \stackrel0: P[\cCons(h,t)/x]}
  {0\Gamma \vdash \mathrm{recList}_{x.P}\,M\,\{\,\cNil \mapsto N_1; \cCons(h,t;p) \mapsto N_2\,\} \stackrel0: P[M/x]}
\end{mathpar}
$\beta$-equalities for recursion:
\begin{mathpar}
  \inferrule* [right=Eq-List-Rec-Nil]
  {0\Gamma, x : \stackrel0: \ListTy(T) \vdash P~\istype \\
    0\Gamma \vdash N_1 \stackrel0: P[\cNil/x] \\
    0\Gamma, h \stackrel\sigma: T, t \stackrel\sigma: \ListTy(T) \vdash N_2 \stackrel0: P[\cCons(h,t)/x]}
  {0\Gamma \vdash \mathrm{recList}_{x.P}\,\cNil\,\{\,\cNil \mapsto N_1; \cCons(h,t;p) \mapsto N_2\,\} \equiv N_1 \stackrel0: P[\cNil/x]}

  \inferrule* [right=Eq-List-Rec-Cons]
  {0\Gamma, x : \stackrel0: \ListTy(T) \vdash P~\istype \\
    0\Gamma \vdash M_1 \stackrel0: T \\
    0\Gamma \vdash M_2 \stackrel0: \ListTy(T) \\
    0\Gamma \vdash N_1 \stackrel0: P[\cNil/x] \\
    0\Gamma, h \stackrel0: T, t \stackrel0: \ListTy(T) \vdash N_2 \stackrel0: P[\cCons(h,t)/x]}
  {0\Gamma \vdash {\begin{array}[m]{@{}l}
                     \mathrm{recList}_{x.P}\,(\cCons(M_1,M_2))\,\{\,\cNil \mapsto N_1; \cCons(h,t;p) \mapsto N_2\,\}\\
                     \equiv N_2[\begin{array}[t]{@{}l} M_1/h,M_2/t,\\
                                  \mathrm{recList}_{x,p}\,M_2\,\{\cNil \mapsto N_1; \cCons(h,t;p) \mapsto N_2\,\}/p]\end{array} \end{array}} \stackrel0: P[\cCons(M_1,M_2)/x]}
\end{mathpar}

\subsection{Cons-Free Naturals} Formation and introduction:
\begin{mathpar}
  \inferrule*
  {0\Gamma~\isctxt}
  {0\Gamma \vdash \tyNat~\istype}

  \inferrule*
  {\Gamma~\isctxt}
  {\Gamma \vdash \conZero \stackrel0: \tyNat}

  \inferrule*
  {\Gamma \vdash M \stackrel0: \tyNat}
  {\Gamma \vdash \conSucc(M) \stackrel0: \tyNat}
\end{mathpar}
Duplication:
\begin{mathpar}
  \inferrule*
  {\Gamma \vdash M \stackrel\sigma: \tyNat}
  {\Gamma \vdash \mathrm{dupNat}(M) \stackrel\sigma: \tyNat \otimes \tyNat}

  \inferrule*
  {\Gamma \vdash M \stackrel0: \tyNat}
  {\Gamma \vdash \mathrm{dupNat}(M) \equiv (M,M) \stackrel0: \tyNat \otimes \tyNat}
\end{mathpar}
Eliminator:
\begin{displaymath}
  \mprset{flushleft}
  \inferrule*
  {0\Gamma, x \stackrel0: \tyNat \vdash P~\istype \\\\
    \Gamma \vdash M \stackrel\sigma: \tyNat \\\\
    0\Gamma \vdash N_z \stackrel\sigma: P[\conZero/x] \\\\
    0\Gamma, n \stackrel0: \tyNat, p \stackrel\sigma: P[n/x] \vdash N_s \stackrel\sigma: P[\conSucc(n)/x]}
  {\Gamma \vdash \tmRec_{x.P}\,M\,\{\conZero \mapsto N_z; \conSucc(n;p) \mapsto N_s\} \stackrel\sigma: P[M/x]}
\end{displaymath}
$\beta$-equalties (only available in the $\sigma = 0$ fragment):
\begin{mathpar}
  \mprset{flushleft}
  \inferrule*
  {0\Gamma, x \stackrel0: \tyNat \vdash P~\istype \\\\
    0\Gamma \vdash N_z \stackrel0: P[\conZero/x] \\\\
    0\Gamma, n \stackrel0: \tyNat, p \stackrel0: P[n/x] \vdash N_s \stackrel0: P[\conSucc(n)/x]}
  {0\Gamma \vdash \tmRec_{x.P}\,\conZero\,\{\conZero \mapsto N_z; \conSucc(n;p) \mapsto N_s\} \equiv N_z \stackrel0: P[\conZero/x]}

  \mprset{flushleft}
  \inferrule*
  {0\Gamma, x \stackrel0: \tyNat \vdash P~\istype \\\\
    0\Gamma \vdash M \stackrel0: \tyNat \\\\
    0\Gamma \vdash N_z \stackrel0: P[\conZero/x] \\\\
    0\Gamma, n \stackrel0: \tyNat, p \stackrel0: P[n/x] \vdash N_s \stackrel0: P[\conSucc(n)/x]}
  {0\Gamma \vdash {\begin{array}{@{}l}
                    \tmRec_{x.P}\,(\conSucc(M))\,\{\conZero \mapsto N_z; \conSucc(n;p) \mapsto N_s\} \\
                    \equiv N_s[M/n,\tmRec_{x.P}\,M\,\{\conZero \mapsto N_z; \conSucc(n;p) \mapsto N_s\}/p] \end{array}} \stackrel0: P[\conSucc(M)/x]}
\end{mathpar}

\subsection{LFPL Diamonds} Formation, introduction, and $\eta$-law:
\begin{mathpar}
  \inferrule*
  {\Gamma~\isctxt}
  {0\Gamma \vdash \Diamond~\istype}

  \inferrule*
  {\Gamma~\isctxt}
  {0\Gamma \vdash * \stackrel0: \Diamond}

  \inferrule*
  {\Gamma \vdash M \stackrel0: \Diamond}
  {\Gamma \vdash M \equiv * \stackrel0: \Diamond}
\end{mathpar}

\subsection{LFPL Naturals} Type formation and introduction:
\begin{mathpar}
  \inferrule*
  {0\Gamma~\isctxt}
  {0\Gamma \vdash \tyNat~\istype}

  \inferrule*
  {\Gamma \vdash M \stackrel\sigma: \Diamond}
  {\Gamma \vdash \conZero(M) \stackrel\sigma: \tyNat}

  \inferrule*
  {\Gamma_1 \vdash M \stackrel\sigma: \Diamond \\
    \Gamma_2 \vdash N \stackrel\sigma: \tyNat \\
    0\Gamma_1 = 0\Gamma_2}
  {\Gamma_1 + \Gamma_2 \vdash \conSucc(M,N) \stackrel\sigma: \tyNat}
\end{mathpar}
Elimination:
\begin{mathpar}
  \mprset{flushleft}
  \inferrule*
  {0\Gamma, x \stackrel0: \tyNat \vdash P~\istype \\\\
    \Gamma \vdash M \stackrel\sigma: \tyNat \\\\
    0\Gamma, d \stackrel\sigma: \Diamond \vdash N_z \stackrel\sigma: P[\conZero(*)/x] \\\\
    0\Gamma, d \stackrel\sigma: \Diamond, n \stackrel0: \tyNat, p \stackrel\sigma: P[n/x] \vdash N_s \stackrel\sigma : P[\conSucc(*,n)/x]}
  {\Gamma \vdash \tmRec\,M\,\{\conZero(d) \mapsto N_z; \conSucc(d,n;p) \mapsto N_s\} \stackrel\sigma: P[M/x]}
\end{mathpar}
$\beta$-equalities, available in both fragments:
\begin{mathpar}
  \mprset{flushleft}
  \inferrule*
  {0\Gamma, x \stackrel0: \tyNat \vdash P~\istype \\\\
    \Gamma \vdash M \stackrel\sigma: \Diamond \\\\
    0\Gamma, d \stackrel\sigma: \Diamond \vdash N_z \stackrel\sigma: P[\conZero(*)/x] \\\\
    0\Gamma, d \stackrel\sigma: \Diamond, n \stackrel0: \tyNat, p \stackrel\sigma: P[n/x] \vdash N_s \stackrel\sigma : P[\conSucc(*,n)/x]}
  {\Gamma \vdash \tmRec\,\conZero(M)\,\{\conZero(d) \mapsto N_z; \conSucc(d,n;p) \mapsto N_s\} \equiv N_z[M/d] \stackrel\sigma: P[\conZero(*)/x]}

  \mprset{flushleft}
  \inferrule*
  {0\Gamma, x \stackrel0: \tyNat \vdash P~\istype \\\\
    \Gamma_1 \vdash M_d \stackrel\sigma: \Diamond \\\\
    \Gamma_2 \vdash M_n \stackrel\sigma: \tyNat \\\\
    0\Gamma, d \stackrel\sigma: \Diamond \vdash N_z \stackrel\sigma: P[\conZero(*)/x] \\\\
    0\Gamma, d \stackrel\sigma: \Diamond, n \stackrel0: \tyNat, p \stackrel\sigma: P[n/x] \vdash N_s \stackrel\sigma : P[\conSucc(*,n)/x] \\\\
    0\Gamma_1 = 0\Gamma_2 = 0\Gamma}
  {\Gamma_1 + \Gamma_2 \vdash {\begin{array}{@{}l}
                                 \tmRec\,(\conSucc(M_d,M_n))\,\{\conZero(d) \mapsto N_z; \conSucc(d,n;p) \mapsto N_s\} \\
                                 \equiv N_s[M_d/d, M_n/n, \tmRec\,M_n\,\{\conZero(d) \mapsto N_z; \conSucc(d,n;p) \mapsto N_s\}/p] \end{array}} \stackrel\sigma: P[\conSucc(*,M_n)/x]}
\end{mathpar}

\subsection{Realisability Reflection} Type formation, introduction and
elimination:
\begin{mathpar}
  \inferrule*
  {0\Gamma \vdash A~\istype}
  {0\Gamma \vdash \Rtype(A)~\istype}

  \inferrule*
  {0\Gamma \vdash M \stackrel1: A}
  {0\Gamma \vdash \rIntro(M) \stackrel\sigma: \Rtype(A)}

  \inferrule*
  {\Gamma \vdash M \stackrel\sigma: \Rtype(A)}
  {\Gamma \vdash \rElim(M) \stackrel{\sigma'}: A}
\end{mathpar}
Equalities:
\begin{mathpar}
  \inferrule*
  {0\Gamma \vdash M \stackrel1: A}
  {0\Gamma \vdash \rElim(\rIntro(M)) \equiv M \stackrel\sigma: A}

  \inferrule*
  {0\Gamma \vdash M \stackrel\sigma: \Rtype(A)}
  {0\Gamma \vdash \rIntro(\rElim(M)) \equiv M \stackrel\sigma: A}
\end{mathpar}